\documentclass[journal]{IEEEtran}
\usepackage{amsmath, amssymb, amsfonts}
\usepackage{bm, lipsum}
\usepackage{physics}
\usepackage{graphicx, enumitem}
\usepackage{color, xcolor, soul, framed}
\usepackage{cancel}
\usepackage{booktabs}
\usepackage{geometry}
\usepackage{hyperref}
\usepackage{mathtools}
\usepackage{cite}
\usepackage{stfloats}
\usepackage{array}
\usepackage{makecell}
\usepackage{subcaption}
\usepackage{booktabs}
\usepackage{mathtools}
\usepackage{varwidth}
\usepackage{multirow}
\usepackage{lipsum}
\usepackage{orcidlink}

\newtheorem{corollary}{Corollary}

\usepackage{diagbox}

\usepackage{framed}
\newtheorem{remark}{Remark}

\usepackage{color, colortbl}
\definecolor{LightCyan}{rgb}{0.88,1,1}
\definecolor{Gray}{gray}{0.93}

\newcommand{\CC}{\cellcolor{Gray!80!white}}
\newcommand{\CB}{\cellcolor{cyan!20!white}}
\newcommand{\CO}{\cellcolor{orange!20!white}}
\newcommand{\CY}{\cellcolor{yellow!30!white}}

\usepackage{tikz}
\usetikzlibrary{arrows.meta, positioning, calc, shapes.geometric}
\definecolor{silver}{HTML}{C0C0C0}
\definecolor{gold}{HTML}{FFD700}

\definecolor{colorFF}{HTML}{F59827}
\definecolor{colorFBe}{HTML}{1565C0}
\definecolor{colorFBx}{HTML}{5EF527}
\definecolor{colorRes}{HTML}{7D7B28}

\hypersetup{
    colorlinks=true,    citecolor=blue,    linkcolor=red,    filecolor=magenta,  urlcolor=blue,    pdftitle={Overleaf Example}}
    
\usepackage{fancyhdr}
\begin{document}

\title{Estimation-Aware Control on Quadrotor State Manifolds via Anisotropic Covariance Filtering}
\date{\today}

\author{Daniel~Engelsman, and~Itzik~Klein
\thanks{The authors are with the Hatter Department of Marine Technologies, Charney School of Marine Sciences, University of Haifa, Israel.\\ E-mails: \{dengelsm@campus, kitzik@univ\}.haifa.ac.il}}

\maketitle
\begin{abstract}
Conventional nonlinear flight controllers operate under the implicit assumption of reliable, high-confidence state feedback. During transient sensing degradation or partial unobservability, state estimation uncertainty is frequently misinterpreted as physical tracking error, triggering aggressive and parasitic corrective inputs. To mitigate this vulnerability, we propose a generalized Estimation-Aware (EA) control framework that directly incorporates real-time estimator confidence into control execution. By mapping the posterior error covariance $\mathbf{P}_{k\mid k}$ into an uncertainty-dependent modulation operator, the EA mechanism dynamically scales feedback authority—performing anisotropic attenuation along weakly observed state directions to prevent false error corrections. Validated across six distinct control paradigms, the proposed approach demonstrates substantial performance gains across all baselines. In particular, when integrated into an Incremental Nonlinear Dynamic Inversion (INDI) architecture, the EA mechanism reduces tracking and estimation errors by 24.34\% and 33.24\%, respectively, at a high flight speed of 10 m/s.
\end{abstract}

\section{Introduction}
\IEEEPARstart{I}{n} modern autonomous robotics, high-agility guidance, navigation, and control (GNC) architectures rely heavily on a foundational assumption: the Separation Principle \cite{slotine1991applied}. Under this paradigm, state estimation and feedback control are synthesized as isolated modules, where the controller treats incoming state estimates as deterministic ground truth \cite{kendoul2012survey}. While this separation holds under nominal sensing conditions, real-world flight environments frequently violate it \cite{HESPANHA19991}. Perception pipelines—such as vision-aided or sensor fusion algorithms—inevitably experience transient degradation due to feature loss, sensor saturation, unmodeled vehicle kinematics, or temporary state unobservability \cite{SIMANDL20011703}.
\\
When state estimation degrades, this classical modular separation introduces a severe structural vulnerability. Standard nonlinear flight controllers evaluate trajectory tracking errors by directly comparing desired reference states against instantaneous state estimates \cite{Jing2020}. During periods of sensor impairments, such as vision loss or sensor saturation, estimator drift and high-variance noise are misidentified by the feedback law as genuine physical tracking errors \cite{isidori1985nonlinear}. Consequently, the controller commands aggressive, high-frequency actuation to correct these phantom deviations \cite{Molin2013}. 
This destructive interaction, rapidly depletes battery energy, excites unmodeled structural dynamics, and further corrupts the state estimator, ultimately culminating in severe kinematic drift \cite{Ryan2007}. To mitigate that, conventional approaches often resort to conservative gain tuning or gain-scheduling schemes \cite{Hongwei2021}. However, such remedies degrade tracking responsiveness and maneuverability during nominal flight merely to safeguard against uncertainty spikes \cite{Dong2021}. Conversely, while stochastic paradigms explicitly account for state estimation uncertainty, their dependence on real-time optimization imposes prohibitive computational overhead for embedded execution on resource-constrained flight microcontrollers \cite{idrissi2022review, KLEIN2025104077, Engelsman2023}.
\\
To address this challenge, we introduce the Estimation-Aware (EA) control framework to bridge state uncertainty and control execution. Rather than assuming deterministic estimates, the EA mechanism explicitly couples feedback authority to the posterior error covariance matrix $\mathbf{P}$. As dynamic excitation restores local observability, the small eigenvalues $\lambda_{\min}(\mathbf{P})$ shrink, allowing high-confidence modal information to reinforce the corresponding controlled states. Concurrently, the associated eigenvectors $\boldsymbol{v}_{i}$ define principal uncertainty directions, suppressing phantom error corrections along unobservable axes.
\begin{figure}[!t] 
\centering 
\begin{tikzpicture}[
    >=Stealth,
    font=\sffamily\tiny,
    cellbox/.style={line width=1.2pt, rectangle, minimum width=4.2cm, minimum height=3.8cm, inner sep=0pt},
    vecFF/.style={-{Stealth[length=7pt, width=6pt]}, densely dashed, line width=2.8pt, cyan!95!white},
    vecFBx/.style={-{Stealth[length=7pt, width=6pt]}, densely dashed, line width=2.8pt, orange!90!white},
    vecFBe/.style={-{Stealth[length=7pt, width=6pt]}, densely dashed, line width=2.8pt, green!60!black},
    vecRes/.style={-{Stealth[length=7pt, width=9pt]}, double, line width=1.8pt, gold!95!black},
    d_line/.style={line width=2.5pt, densely dashed, gray!70!white},
]

  \node[cellbox] (cell1) at (0,0) {};
  \node[cellbox] (cell2) at (4.5,0) {};
  
  \begin{scope}[shift={(cell1.center)}]
    \path let \p1 = ($(cell1.north east)-(cell1.south west)$) in
    node[anchor=center, inner sep=0pt, opacity=0.8] at (0,0) {%
      \includegraphics[width=\x1, height=\y1]{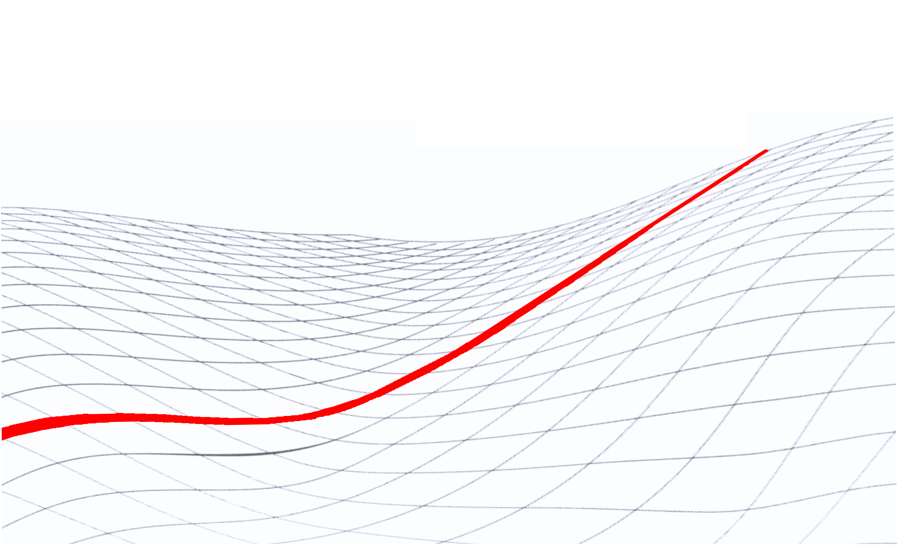}%
    };

    \node[anchor=north, font=\large] at (-1.8, 1.9) {($i$)};
    
    \coordinate (O1) at (-1.3, -0.0);
    \draw[black, double, line width=1.5pt, line cap=round] 
    ($(O1)+(-9pt,-5pt)$) -- ($(O1)+(9pt,5pt)$)
    ($(O1)+(-5pt,7pt)$)  -- ($(O1)+(5pt,-7pt)$);

    \draw[vecFF] (O1) -- ++(30:1.1) coordinate (FF1);
    \draw[vecFBe] (O1) -- ++(-80:1.05) coordinate (FBe1);
    \coordinate (R1) at ($(FF1) + (FBe1) - (O1) + (5:-0.15)$);
    \draw[vecRes] (O1) -- (R1);

    \draw[d_line] plot [smooth, tension=1.0] coordinates {
    ($(O1)+(1.0, -.5)$) 
    ($(O1)+(1.72, -.72)$) 
    ($(O1)+(1.3, -0.35)$)
    ($(O1)+(1.85, -0.3)$)
    ($(O1)+(1.75, -0.05)$)
    ($(O1)+(2.3, 0.35)$)
    ($(O1)+(2.8, 0.85)$)
    };
    
    \draw[black, line width=0.8pt] (cell1.south west) rectangle (cell1.north east);
  \end{scope}

  \begin{scope}[shift={(cell2.center)}]
    \path let \p1 = ($(cell1.north east)-(cell1.south west)$) in
    node[anchor=center, inner sep=0pt, opacity=0.8] at (0,0) {%
      \includegraphics[width=\x1, height=\y1]{Photos/Img_Grid_Top.png}%
    };
    
    \node[anchor=north, font=\large] at (-1.75, 1.9) {($ii$)};
    
    \coordinate (O2) at (-1.3, -0.0);
    \draw[black, double, line width=1.5pt, line cap=round] 
    ($(O2)+(-9pt,-5pt)$) -- ($(O2)+(9pt,5pt)$)
    ($(O2)+(-5pt,7pt)$)  -- ($(O2)+(5pt,-7pt)$);

    \draw[vecFF]  (O2) -- ++(30:1.1) coordinate (FF2);
    \draw[vecFBe] (O2) -- ++(-80:1.05) coordinate (FBe2);
    \draw[vecFBx] (O2) -- ++(100:0.9) coordinate (FBx2);
    \coordinate (R2) at ($(O2) + ($(FF2)-(O2)$) + ($(FBe2)-(O2)$) + ($(FBx2)-(O2)$) + ($(-95:0.4)$)$);
    \draw[vecRes] (O2) -- (R2);

    \draw[d_line] plot [smooth, tension=0.5] coordinates {
    ($(O2)+(0.95, .0)$) 
    ($(O2)+(1.8, .08)$)
    ($(O2)+(2.8, 0.85)$)
    };
    
    \draw[black, line width=0.8pt] (cell2.south west) rectangle (cell2.north east);
  \end{scope}
\end{tikzpicture}
\caption{(i) Standard dual-loop control (FB+FF) tends to react aggressively to spatial errors, inducing oscillatory hunting across the reference path. 
(ii) The proposed augmentation (EA) tempers immediate error correction to damp rate transients, sacrificing instant recovery for long-term stability.}
\label{fig:diagram_1}
\end{figure}
%
Fig.~\ref{fig:diagram_1} illustrates this physical trade-off on the state space manifold. In standard dual-loop tracking (i), spatial feedback (FB, green) and predictive feedforward (FF, cyan) synthesize a net command vector (yellow) that overreacts to perceived errors, inducing oscillatory hunting around the reference path (red curve). Conversely, with the proposed EA mechanism (ii), control effort prioritizes long-term trajectory convergence over instant spatial correction. 
By leveraging inner-loop rate damping (orange) to temper immediate restoration, the net command vector achieves smooth, asymptotic convergence to the near-future reference trajectory.
\\
The primary contributions of this work are three-fold:
\begin{enumerate}[label=(\roman*)]
    \item \textbf{Estimation-Aware Control Operator:} We formulate an anisotropic, uncertainty-gated operator driven by posterior error covariance, providing dynamic attenuation of feedback authority without requiring ad-hoc gain tuning.
    \item \textbf{Parasitic Actuation Mitigation:} We establish a control-theoretic mechanism that explicitly prevents state estimator uncertainty from being misidentified as physical tracking error, eliminating high-frequency control chatter and self-excitation during transient sensor outages.
    \item \textbf{Broad Controller Integration:} We systematically integrate and evaluate the proposed mechanism across six prominent quadrotor control architectures, demonstrating its model-agnostic flexibility.
\end{enumerate}

In extensive trials featuring severe estimation degradation, the proposed method reduces kinematic tracking drift by 24.34\% and state estimation error by 33.24\%. 
\\
The remainder of this paper is organized as follows: Section~\ref{sec:theory} provides the theoretical background, Section~\ref{sec:prob} establishes the problem formulation, and Section~\ref{sec:method} details the proposed methodology. Finally, Section~\ref{sec:results} presents numerical and experimental evaluations, and Section~\ref{sec:conc} concludes the paper.
\section*{Supplementary Materials}
The implementation and supplementary materials are publicly available at
the project repository @ \href{https://github.com/ansfl/Estimation-Aware-Control}{\textbf{\texttt{GitHub}}}.

\section{Preliminaries} \label{sec:theory}
This section outlines the theoretical preliminaries for nonlinear tracking control. Specifically, it reviews the physical system dynamics, feedback linearization principles, and state estimation frameworks that underpin the proposed methodology.

\subsection{Nonlinear Dynamical System}
We consider a class of discrete-time, control-affine systems obtained by sampling the continuous-time Newton-Euler rigid-body dynamics over a constant interval $\Delta t$. Letting $k \in \mathbb{N}$ denote the discrete time index, the system dynamics evolve according to \cite{bar2001estimation}
\begin{align} \label{eq:state_space_GT}
\mathcal{S} : \left\{
\begin{aligned}
\boldsymbol{x}_{k+1} &= \boldsymbol{f}(\boldsymbol{x}_k) + \mathbf{g}(\boldsymbol{x}_k)\boldsymbol{u}_k + \boldsymbol{w}_k \, , 
\\
\boldsymbol{y}_k &= \boldsymbol{h}(\boldsymbol{x}_k) + \boldsymbol{v}_k \, ,
\end{aligned} \right.
\end{align}
where $k \in \mathbb{N}_0$ denotes the discrete time index, $\boldsymbol{x}_k \in \mathcal{M}$ represents the state vector on an $n$-dimensional smooth manifold $\mathcal{M}$, $\boldsymbol{u}_k \in \mathbb{R}^{m}$ is the control input vector, and $\boldsymbol{y}_k \in \mathbb{R}^{p}$ is the vector of measurable outputs. The mapping of $\boldsymbol{f} : \mathcal{M} \to \mathcal{M}$ characterizes the discrete unforced system propagation, $\mathbf{g} : \mathcal{M} \to \mathbb{R}^{n \times m}$ is the input distribution matrix, and $\boldsymbol{h} : \mathcal{M} \to \mathbb{R}^p$ denotes the observation model. The process disturbance $\boldsymbol{w}_k \in \mathbb{R}^{n}$ and measurement noise $\boldsymbol{v}_k \in \mathbb{R}^{p}$ are modeled as mutually uncorrelated, zero-mean discrete Gaussian white noise processes with covariances $\boldsymbol{\Sigma}_{w,k} \succeq 0$ and $\boldsymbol{\Sigma}_{v,k} \succ 0$, respectively.
\\
\subsection{Nonlinear Observability and Metric Conditioning} \label{subsec:obs}
Observability establishes whether the state trajectory can be uniquely reconstructed from input-output histories. For the nonlinear system \eqref{eq:state_space_GT}, local weak observability is characterized by the codistribution $\Omega_{\mathrm{NL}}(\boldsymbol{x}) \subset T^*_{\boldsymbol{x}}\mathcal{M}$, defined as the span of the differentials of the output function and its successive Lie derivatives along the system vector field $\boldsymbol{f}$ \cite{martinelli2018nonlinear}
\begin{align} \label{eq:codistribution}
\Omega_{\mathrm{NL}}(\boldsymbol{x}_k) \triangleq \mathrm{span}\left\{ \mathrm{d}L_f^k h_i(\boldsymbol{x}) \mid k \ge 0 \right\} .
\end{align}
Evaluating the local gradients yields the nonlinear observability matrix $\mathcal{O}_{\mathrm{NL}}(\boldsymbol{x}_k) \in \mathbb{R}^{pn \times n}$:
\begin{align} \label{eq:obs_NL}
\mathcal{O}_{\mathrm{NL}}(\boldsymbol{x}_k) \triangleq \begin{pmatrix}
\frac{\partial \boldsymbol{h}(\boldsymbol{x}_k)}{\partial \boldsymbol{x}_k}^\top & 
\dots & 
\frac{\partial L_f^{n-1} \boldsymbol{h}(\boldsymbol{x}_k)}{\partial \boldsymbol{x}_k}^\top
\end{pmatrix}^\top .
\end{align}
If $\mathrm{rank}\left(\mathcal{O}_{\mathrm{NL}}(\boldsymbol{x}_k)\right) = n$, the system satisfies the Hermann-Krener local observability rank condition at $\boldsymbol{x}_k$; otherwise, unobserved state modes emerge \cite{Hermann1977}.
Quantifying observability strength along a trajectory $\boldsymbol{x}(t)$ over $[t_0, t_f]$, elevates the local observability Gramian $\boldsymbol{W}_{\mathrm{O}}(t_0, t_f) \in \mathbb{R}^{n \times n}$:
\begin{align} \label{eq:Gramian}
\boldsymbol{W}_{\mathrm{O}}(t_0, t_f) \triangleq \int_{t_0}^{t_f} \boldsymbol{\Phi}^\top(\tau, t_0) \, \mathbf{H}^\top(\tau) \, \mathbf{H}(\tau) \, \boldsymbol{\Phi}(\tau, t_0) \, \mathrm{d}\tau,
\end{align}
where $\mathbf{H}(\tau) \triangleq \left. \frac{\partial \boldsymbol{h}}{\partial \boldsymbol{x}} \right|_{\boldsymbol{x}(\tau)}$ is the measurement Jacobian and $\boldsymbol{\Phi}(\tau, t_0)$ governs the error system dynamics \eqref{eq:x_tilde}. While rank sufficiency guarantees observability, the condition number 
\begin{align} \label{eq:condition}
\kappa(\boldsymbol{W}_{\mathrm{O}}) = \lambda_{\max}(\boldsymbol{W}_{\mathrm{O}}) / \lambda_{\min}(\boldsymbol{W}_{\mathrm{O}})\, ,
\end{align}
dictates directional state uncertainty, where large values signal severe ill-conditioning along weakly excited axes \cite{Gauthier1994}.
%
%
%

\subsection{Feedforward-Feedback Dynamic Inversion}
A control problem seeks an input sequence $\boldsymbol{u}_k \in \mathbb{R}^m$ to steer a system state $\boldsymbol{x}_k \in \mathbb{R}^n$ toward a reference state $\boldsymbol{x}_{\mathrm{ref},k}$. The full-state error is defined as
\begin{align} \label{eq:epsilon_x}
\boldsymbol{\epsilon}_{\boldsymbol{x},k} = \boldsymbol{x}_k - \boldsymbol{x}_{\mathrm{ref},k} \, \in \, \mathbb{R}^{n} \, ,
\end{align}
corresponding to a regulation task if the reference is time-invariant  $\dot{\boldsymbol{x}}_{\mathrm{ref}} = \mathbf{0}$, or a tracking task, $\dot{\boldsymbol{x}}_{\mathrm{ref}}\neq\boldsymbol{0}$, otherwise.
In practical robotic platforms, system are mostly underactuated and partially observed ($p \le m < n$), rendering full-state feedback unfeasible. Control design is thus formulated over a set of task-relevant controlled (flat) outputs, yielding the practical output tracking error
\begin{align} \label{eq:epsilon_y}
\boldsymbol{\epsilon}_{\boldsymbol{y},k} = \boldsymbol{y}_k - \boldsymbol{y}_{\mathrm{ref},k} \in \mathbb{R}^{p} \, .
\end{align}
For a discrete system with vector relative degree $r$, the control input $\boldsymbol{u}_k$ explicitly affects the output at the $r$-th forward discrete time step \cite{acquatella2012robust}
\begin{align} \label{eq:normal_form_discrete}
\boldsymbol{y}_{k+r} = \boldsymbol{\alpha}(\boldsymbol{x}_k) + \boldsymbol{\beta}(\boldsymbol{x}_k)\boldsymbol{u}_k \, ,
\end{align}
where $\boldsymbol{\alpha}(\boldsymbol{x}_k) = \mathcal{L}_{\boldsymbol{f}}^r \boldsymbol{h}(\boldsymbol{x}_k)$ represents the unforced $r$-step output propagation, and $\boldsymbol{\beta}(\boldsymbol{x}_k) = \frac{\partial (\boldsymbol{h} \circ \boldsymbol{f}^{r-1})}{\partial \boldsymbol{x}_k}\mathbf{G}(\boldsymbol{x}_k)$ denotes the decoupling matrix. Standard nonlinear dynamic inversion (NDI) enforces a desired output evolution $\boldsymbol{\nu}_k \triangleq \boldsymbol{y}_{k+r}$ via
\begin{align} \label{eq:control_law_NDI}
\boldsymbol{u}_k = \boldsymbol{\beta}^\dagger(\boldsymbol{x}_k) \bigl( \boldsymbol{\nu}_k - \boldsymbol{\alpha}(\boldsymbol{x}_k) \bigr) \, ,
\end{align}
where $\boldsymbol{\beta}^\dagger = \boldsymbol{\beta}^\top (\boldsymbol{\beta}\boldsymbol{\beta}^\top)^{-1} \in \mathbb{R}^{m \times p}$ denotes the right pseudoinverse \cite{smeur2018cascaded, Bacon2000control}.
\newpage
To guarantee exponential tracking error decay, the output state space across the relative degree window $\{\boldsymbol{\epsilon}_{\boldsymbol{y},k}, \, \boldsymbol{\epsilon}_{\boldsymbol{y},k+1}, \, \dots, \, \boldsymbol{\epsilon}_{\boldsymbol{y},k+r-1}\}$ is constrained to follow the linear error difference equation
\begin{align} \label{eq:error_diff_eq}
\boldsymbol{\epsilon}_{\boldsymbol{y},k+r} + \sum\nolimits_{j=0}^{r-1} \mathbf{K}_j \boldsymbol{\epsilon}_{\boldsymbol{y},k+j} = \mathbf{0} \, ,
\end{align}
where gain matrices $\mathbf{K}_j \in \mathbb{R}^{p \times p}$ are selected such that the characteristic error polynomial is Schur stable \cite{tal2020accurate}. 

Under exact model knowledge and uncorrupted state feedback, substituting \eqref{eq:control_law_NDI} into \eqref{eq:normal_form_discrete} cancels system nonlinearities, yielding the decoupled $r$-step integrator dynamics
\begin{align} \label{eq:decoupled_integrator}
\boldsymbol{y}_{k+r} = \boldsymbol{\nu}_k \, .
\end{align}
Lastly, substituting $\boldsymbol{\epsilon}_{\boldsymbol{y},k+r} = \boldsymbol{y}_{k+r} - \boldsymbol{y}_{\mathrm{ref},k+r}$ into \eqref{eq:error_diff_eq} alongside \eqref{eq:decoupled_integrator} yields the discrete virtual control command
\begin{align} \label{eq:y_dot_ref}
\boldsymbol{\nu}_k = \underbrace{ \boldsymbol{y}_{ \mathrm{ref},k+r}\vphantom{\sum\nolimits_{j=0}^{r-1}} }_{\mathrm{FF}} - \underbrace{ \sum\nolimits_{j=0}^{r-1}  \mathbf{K}_j \boldsymbol{\epsilon}_{\boldsymbol{y},k+j} }_{\mathrm{FB, outer}} \, .
\end{align}
\begin{remark}[Geometric Interpretation] \label{rem:geometric_decomp}
Let $\mathbf{t}_k \in \mathrm{T}_{\boldsymbol{x}}\mathcal{M}$ denote the local unit tangent vector aligned with desired trajectory progression $\boldsymbol{y}_{\mathrm{ref}}$. As shown, \eqref{eq:y_dot_ref} decouples within $\mathrm{T}_{\boldsymbol{x}}\mathcal{M}$ into predictive feedforward, acting tangentially along the horizon $r$ steps ahead ($\mathrm{FF}_k \parallel \mathbf{t}_{k+r}$), and feedback (FB) correction, providing immediate normal restoration ($\mathrm{FB}_k \perp \mathbf{t}_k$). 
\end{remark}

\subsection{Nonlinear State Estimation}
Since the true system state $\boldsymbol{x}_k$ remains latent, an estimate $\hat{\boldsymbol{x}}_k$ is constructed from available inputs $\boldsymbol{u}_k$ and measurements $\boldsymbol{y}_k$. Following the system structure \eqref{eq:state_space_GT}, a nonlinear observer is formulated in discrete predictor--corrector form \cite{kay1993fundamentals}:
\begin{align} \label{eq:state_space_est}
\widehat{\mathcal{S}} : \left\{ 
\begin{aligned}
\hat{\boldsymbol{x}}_{k+1}^- &= \boldsymbol{f}(\hat{\boldsymbol{x}}_k^+) + \mathbf{g}(\hat{\boldsymbol{x}}_k^+) \boldsymbol{u}_k \, , \\
\hat{\boldsymbol{x}}_k^+ &= \hat{\boldsymbol{x}}_k^- + \boldsymbol{\ell}(\hat{\boldsymbol{x}}_k^-, \boldsymbol{y}_k) \, , \\
\hat{\boldsymbol{y}}_k &= \boldsymbol{h}(\hat{\boldsymbol{x}}_k^-) \, ,
\end{aligned} \right.
\end{align}
where superscripts $(\cdot)^-$ and $(\cdot)^+$ denote the a priori (predicted) and a posteriori (corrected) state estimates, respectively. The prediction phase propagates the state along the unforced map $\boldsymbol{f}(\cdot)$ and input matrix $\mathbf{g}(\cdot)$, while the correction phase employs an innovation term $\boldsymbol{\ell}(\cdot, \cdot)$ to inject measurement output residuals into the state space. 
\\
The corresponding state and output estimation errors are \cite{bar2001estimation}
\begin{align}
\tilde{\boldsymbol{x}}_k^\pm &\triangleq \boldsymbol{x}_k - \hat{\boldsymbol{x}}_k^\pm \, , \label{eq:x_tilde} \\
\tilde{\boldsymbol{y}}_k &\triangleq \boldsymbol{y}_k - \hat{\boldsymbol{y}}_k \, . \label{eq:output_inn}
\end{align}

\paragraph{Linearized Stochastic Estimation}
Local tangent estimators linearize the system dynamics and measurement map about the current state estimate \cite{Bacon2000control}. For control-affine systems, the state and measurement Jacobians are:
\begin{align} \label{eq:jacobians}
\mathbf{F}_{\hat{\boldsymbol{x}},k} \triangleq \left. \frac{\partial \left[\boldsymbol{f}(\boldsymbol{x}) + \mathbf{g}(\boldsymbol{x})\boldsymbol{u}_k\right]}{\partial \boldsymbol{x}} \right|_{\hat{\boldsymbol{x}}_k^+} , \quad
\mathbf{H}_{\hat{\boldsymbol{x}},k} \triangleq \left. \frac{\partial \boldsymbol{h}}{\partial \boldsymbol{x}} \right|_{\hat{\boldsymbol{x}}_k^-} .
\end{align}
Assuming deterministic control inputs ($\tilde{\boldsymbol{u}}_k = \mathbf{0}$), first-order Taylor expansions approximate the physical system as:
\begin{align}
\boldsymbol{x}_{k+1} &= \boldsymbol{f}(\hat{\boldsymbol{x}}_k^+) + \mathbf{g}(\hat{\boldsymbol{x}}_k^+)\boldsymbol{u}_k + \mathbf{F}_{\hat{\boldsymbol{x}},k}\tilde{\boldsymbol{x}}_k^+ + \boldsymbol{w}_k + \mathcal{O}\left(\|\tilde{\boldsymbol{x}}_k^+\|^2\right) , \\
\boldsymbol{y}_k &= \boldsymbol{h}(\hat{\boldsymbol{x}}_k^-) + \mathbf{H}_{\hat{\boldsymbol{x}},k}\tilde{\boldsymbol{x}}_k^- + \boldsymbol{v}_k + \mathcal{O}\left(\|\tilde{\boldsymbol{x}}_k^-\|^2\right) .
\end{align}
For a consistent observer, the correction satisfies $\boldsymbol{\ell}\bigl(\hat{\boldsymbol{x}}_k^-, \boldsymbol{h}(\hat{\boldsymbol{x}}_k^-)\bigr) = \mathbf{0}$. Evaluating the first-order expansion of $\boldsymbol{\ell}$ with respect to the output innovation \eqref{eq:output_inn} yields:
\begin{align} \label{eq:observer_approx}
\boldsymbol{\ell}(\hat{\boldsymbol{x}}_k^-, \boldsymbol{y}_k) = \mathbf{L}_{\hat{\boldsymbol{x}},k} \tilde{\boldsymbol{y}}_k + \mathcal{O}\left(\|\tilde{\boldsymbol{y}}_k\|^2\right) ,
\end{align}
where $\mathbf{L}_{\hat{\boldsymbol{x}},k}$ is the state-dependent gain matrix (e.g., the extended Kalman filter gain).

\begin{figure}[h]
\centering
\begin{tikzpicture}[>=Latex,
node distance=16mm and 24mm,
block/.style = {draw, rounded corners=3pt, thick, align=center, inner sep=4pt, minimum width=14mm, minimum height=10mm},
labelnode/.style={font=\normalsize, inner sep=1pt},
every node/.style={font=\normalsize},  
line_ax/.style={-{Stealth[length=5pt, width=5pt]}, line width=1.65pt},
line_er/.style={<->, >={Stealth[length=4pt, width=4pt]}, line width=0.8pt},
line_f_r/.style={red!85!white, line width=3.pt, dotted},
line_df/.style={magenta!80!black, line width=3.2pt, dotted},
line_f_x/.style={cyan!95!white, line width=3.2pt},
d_O/.style={green!90!yellow, line width=3.2pt},
d_line/.style={line width=1.pt, dashed},
frame/.style={line width=4pt, draw=gray!50},
circ/.style={thick, inner sep=0pt},
scale=\linewidth/0.8\textwidth]
]

\def\Lx{0}      \def\Ly{0}
\coordinate (O)     at (\Lx,\Ly);
\draw[line_ax] (O) -- ++(0,6);
\draw[line_ax] (O) -- ++(7.85,0);
\node at ($(O)+(-1.1, 6)$){};
\node at ($(O)+(6, -0.5)$){};

\node[circ] at (3.5, -0.8) (x_true){};
\node[circ] at ($(x_true)+(0, 3.15)$) (f_x_true){};
\node[circ] at ($(x_true)+(2, 5.5)$) (f_x_ref){};
\node[circ] at ($(x_true)+(-2.0, 1.75)$) (f_x_hat){};
\node[circ] at ($(f_x_true)+(0,-0.7)$) (f_x_dyn){};
\node[circ] at ($(O)+(7.95, -0.3)$){$\boldsymbol{\zeta}$};
\node[circ] at ($(O)+(-0.85, 5.75)$){$\boldsymbol{f(\zeta})$};
\node[circ] at ($(O)+(10, 0)$){};

\draw[line_f_x]($(O)+(0.7, .1)$) .. controls ($(f_x_hat)+(0,0)$) and ($(f_x_true)+(-.85,-0.4)$) .. ($(f_x_ref)+(0.6,-2.8)$);
\draw[line_f_r]($(O)+(0.08, 3.5)$) .. controls ($(f_x_hat)+(2,2.2)$) and ($(f_x_true)+(1.2,1.75)$) .. ($(f_x_ref)+(0.6,0.48)$);
\draw[line_df]($(O)+(0.25, 0.12)$) --++ (4.2, 2.9);
\draw[d_O] (f_x_dyn) --++ (f_x_true){}; 

\draw[d_line] (f_x_hat) -- ++(0, -3.3);
\draw[d_line] (f_x_hat) -- ++(8, 0);
\draw[d_line] (f_x_true) -- ++(0, -3.92);
\draw[d_line] (f_x_true) -- ++(4.9, 0);
\draw[d_line] (f_x_ref) -- ++(0, -7.05);
\draw[d_line] (f_x_ref) -- ++(4, 0);

\node[circ, fill=white, fill opacity=1] at (f_x_hat) {\large$\circ$};
\node[circ, fill=white, fill opacity=1] at (f_x_true) {\large$\circ$};
\node[circ, fill=white, fill opacity=1] at (f_x_ref) {\large$\circ$};
\node[circ, fill=white, fill opacity=1] at (f_x_dyn) {\tiny $\boldsymbol{\otimes}$};

\node[fill=white, fill opacity=1, inner sep=2.5pt] at ($(x_true)+(0,0)$) {$\boldsymbol{x}_k$};
\node[fill=white, fill opacity=1, inner sep=2pt] at ($(f_x_hat)+(0, -1.68)$) {$\hat{\boldsymbol{x}}_k$};
\node[fill=white, text opacity=1, fill opacity=1.0] at ($(f_x_ref)+(0.3, -5.55)$) {${\boldsymbol{x}}_{\mathrm{ref},k}$};

\node[fill=white, fill opacity=1, inner sep=1pt] at ($(f_x_hat)+(5.5, 0)$) {$\boldsymbol{f}(\hat{\boldsymbol{x}}_k)$};
\node[fill=white, fill opacity=1, inner sep=1pt] at ($(f_x_true)+(3.5, 0)$) {$\boldsymbol{f}(\boldsymbol{x}_k)$};
\node[fill=white, fill opacity=1, inner sep=1pt] at ($(f_x_ref)+(1.55, -.05)$) {$\boldsymbol{x}_{\mathrm{ref},k+1}$};

\draw[line_er] ($(f_x_hat)+(0.05,-2.5)$) -- ++(2-0.1, 0);
\draw[line_er] ($(f_x_hat)+(2.05,-2.5)$) -- ++(2-0.1, 0);
\draw[line_er] ($(f_x_hat)+(0.05,-3.3)$) -- ++(4-0.05, 0);
\node[circ, fill=white, fill opacity=1] at ($(f_x_hat)+(1.0,-2.43)$) {$\, \tilde{\boldsymbol{x}}_k \, $};
\node[circ, fill=white, fill opacity=1] at ($(f_x_hat)+(3.0,-2.58)$) {$\, \boldsymbol{\epsilon}_{\boldsymbol{x},k} \, $};
\node[circ, fill=white, fill opacity=1, inner sep=2pt] at ($(f_x_hat)+(2.0,-3.2)$) {$\, \hat{\boldsymbol{\epsilon}}_{\boldsymbol{x},k} \, $};

\draw[line_er] ($(f_x_true)+(4.9,0.05)$) -- ++(0.0, 2.22);
\draw[line_er] ($(f_x_hat)+(6.9,0.05)$) -- ++(0.0, 1.32);
\draw[line_er] ($(f_x_hat)+(7.9,0.05)$) -- ++(0.0, 3.6);
\node[circ, fill=white, fill opacity=1, inner sep=1.5pt] at ($(f_x_hat)+(6.9,0.71)$) {$\, \tilde{\boldsymbol{x}}_{k+1} \, $};
\node[circ, fill=white, fill opacity=1, inner sep=1.5pt] at ($(f_x_hat)+(6.9,2.55)$) {$\, \boldsymbol{\epsilon}_{\boldsymbol{x},k+1} \, $};
\node[circ, fill=white, fill opacity=1, inner sep=1.5pt] at ($(f_x_hat)+(8.1,1.85)$) {$\, \hat{\boldsymbol{\epsilon}}_{\boldsymbol{x},k+1} \, $};
\end{tikzpicture}
\caption{Geometric breakdown of reference tracking and state estimation divergence. Linearization of the unforced discrete map $\boldsymbol{f}(\cdot)$ about the estimate $\hat{\boldsymbol{x}}_k$ introduces higher-order model residuals $\mathcal{O}(\|\tilde{\boldsymbol{x}}_k\|^2)$ relative to the true latent dynamics.}
\label{fig:diagram}
\end{figure}
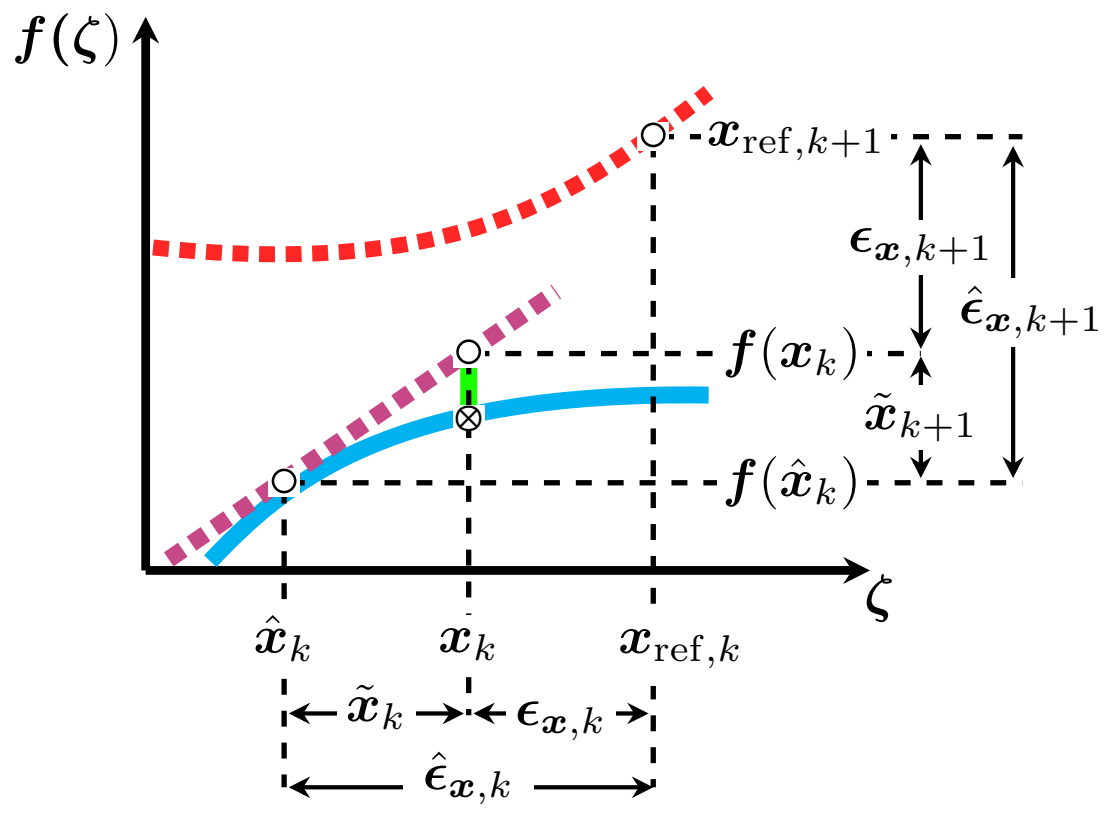

\paragraph{Perturbation Error Dynamics}
The discrepancy between the true state trajectory \eqref{eq:state_space_GT} and the estimated belief \eqref{eq:state_space_est} gives rise to the error operator:
\begin{align} \label{eq:S_residual}
\widetilde{\mathcal{S}} \triangleq \mathcal{S} \ominus \widehat{\mathcal{S}} \, .
\end{align}
Combining the linearized transition maps with \eqref{eq:observer_approx} yields the first-order perturbation error system:
\begin{align} \label{eq:error_space}
\widetilde{\mathcal{S}} : \left\{ 
\begin{aligned}
\tilde{\boldsymbol{x}}_{k+1}^- &= \mathbf{F}_{\hat{\boldsymbol{x}}, k} \tilde{\boldsymbol{x}}_k^+ + \boldsymbol{w}_k \, , \\
\tilde{\boldsymbol{x}}_k^+ &= (\mathbf{I} - \mathbf{L}_{\hat{\boldsymbol{x}},k} \mathbf{H}_{\hat{\boldsymbol{x}}, k}) \tilde{\boldsymbol{x}}_k^- - \mathbf{L}_{\hat{\boldsymbol{x}},k}\boldsymbol{v}_k \, .
\end{aligned} \right.
\end{align}
Unlike linear systems where the separation principle guarantees decoupled controller and observer synthesis \cite{engelsman2026revisiting}, nonlinear estimation and state feedback remain intrinsically coupled through the state-dependent Jacobians $\mathbf{F}_{\hat{\boldsymbol{x}},k}$ and $\mathbf{H}_{\hat{\boldsymbol{x}},k}$, alongside higher-order residual terms \cite{khalil1996robust}.
\\
Fig.~\ref{fig:diagram} illustrates the resulting state divergence, where the true discrete state trajectory $\boldsymbol{x}_k$ (cyan) strives to track the reference trajectory $\boldsymbol{x}_{\mathrm{ref},k}$ (red). This tracking discrepancy is further amplified by estimation uncertainty, as the prior state estimate $\hat{\boldsymbol{x}}_k^-$ (magenta) is propagated at step $k$ via the unforced map $\boldsymbol{f}(\hat{\boldsymbol{x}}_k^+)$. In high-bandwidth operating regimes, linearization residuals (green) intensify as higher-order Taylor terms $\mathcal{O}(\Vert{}\tilde{\boldsymbol{x}}_k\Vert{}^2)$ become non-negligible, causing first-order local approximations to lose fidelity.
\\
In summary, while the state realizations $\boldsymbol{x}_k$ of the plant $\mathcal{S}$ and the estimates $\hat{\boldsymbol{x}}_k^\pm$ of the observer $\widehat{\mathcal{S}}$ evolve on the smooth manifold $\mathcal{M}$, the estimation perturbation vectors $\tilde{\boldsymbol{x}}_k^\pm$ of the error system $\widetilde{\mathcal{S}}$ reside in the local tangent space $T_{\hat{\boldsymbol{x}}_k} \mathcal{M}$. 

\section{Problem Formulation} \label{sec:prob}
Building upon the geometric foundation in Section~\ref{sec:theory}, this section formalizes the structural challenge arising from the cyclic interdependence between tracking error, control policy, and estimation quality under partial state observability.

\paragraph{Bayesian Information Coupling}
From a probabilistic perspective, the latent state $\boldsymbol{x}_k$ is governed by the conditional belief distribution $p(\boldsymbol{x}_k \mid \mathcal{Y}_k)$, where $\mathcal{Y}_k \triangleq \{\boldsymbol{y}_0, \dots, \boldsymbol{y}_k\}$ denotes the historical measurement sequence. Under the local Gaussian assumption, this belief state is fully parameterized by its mean and error covariance
\begin{align} \label{eq:bayesian_belief}
\boldsymbol{x}_k \mid \mathcal{Y}_k \sim \mathcal{N}\left( \hat{\boldsymbol{x}}_k^\pm, \mathbf{P}_k^\pm \right) ,
\end{align}
where $\mathbf{P}_k^\pm \triangleq \mathbb{E}\left[ \tilde{\boldsymbol{x}}_k^\pm (\tilde{\boldsymbol{x}}_k^\pm)^\top \mid \mathcal{Y}_k \right]$ quantifies estimation uncertainty across operational regimes. 
The controller-estimator coupling manifests in the evaluation of the NDI control law \eqref{eq:control_law_NDI} using the perceived belief mean
\begin{align} \label{eq:expected_control}
\boldsymbol{u}_k = \boldsymbol{\kappa}\left(\hat{\boldsymbol{x}}_k^\pm, \boldsymbol{x}_{\mathrm{ref},k}\right) = \boldsymbol{\beta}^\dagger(\hat{\boldsymbol{x}}_k^\pm) \left( \boldsymbol{\nu}_k - \boldsymbol{\alpha}(\hat{\boldsymbol{x}}_k^\pm) \right) .
\end{align}
Taylor expanding the decoupling maps around $\hat{\boldsymbol{x}}_k^\pm$ demonstrates that the expected output tracking error depends explicitly on the trace of the belief covariance $\mathbf{P}_k^\pm$
\begin{align} \label{eq:covariance_degradation}
\mathbb{E}\left[ \boldsymbol{y}_{k+r} - \boldsymbol{\nu}_k \right] \approx \frac{1}{2} \operatorname{tr}\left( \left. \frac{\partial^2 \boldsymbol{\alpha}}{\partial \boldsymbol{x}^2} \right|_{\hat{\boldsymbol{x}}_k^\pm} \mathbf{P}_k^\pm \right) + \mathcal{O}\left( \operatorname{tr}(\mathbf{P}_k^\pm)^2 \right) .
\end{align}

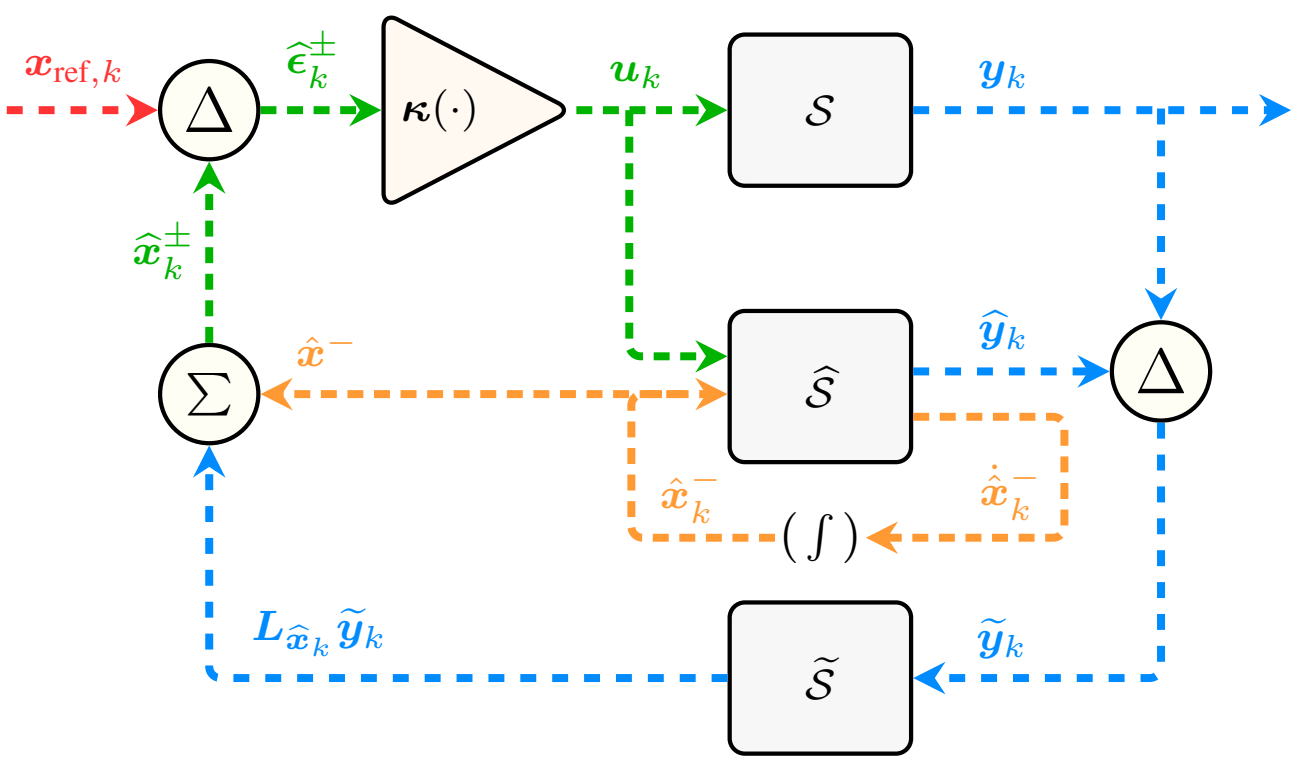
\begin{figure}[b] 
\begin{tikzpicture}[ 
  block/.style={draw, thick, minimum width=12mm, minimum height=1cm, fill=gray!6, rounded corners=1mm},
  joint/.style={draw, fill=yellow!5, circle, thick, minimum size=6.5mm, inner sep=0pt},
  arrow/.style={-{Stealth[length=6pt, width=6pt]}, rounded corners=0.75mm, line width=1.5pt},
  labelnode/.style={font=\footnotesize, inner sep=1pt},
  every node/.style={font=\footnotesize},
  circ/.style={thick, inner sep=0pt},
]

\node[joint] (J_sub_1) {\large $\Delta$};
\node[joint, below=12mm of J_sub_1] (J_sum) {$\sum$};
\node[isosceles triangle, draw, thick, isosceles triangle apex angle=55, rounded corners=1mm, fill=orange!5, minimum width=3mm, minimum height=3mm, right=8mm of J_sub_1] (gain) {$\boldsymbol{\kappa} ( \cdot )$};
\node[block, right=10mm of gain] (SS_GT) {${\mathcal{S}}$};
\node[block, below=8mm of SS_GT] (SS_est) {$\widehat{\mathcal{S}}$};
\node[joint, right=13mm of SS_est, yshift=1mm] (J_sub_2) {\large $\Delta$};
\node[block, below=9mm of SS_est] (SS_dif) {$\widetilde{\mathcal{S}}$};
\node[circ] (J_int) at ($(SS_est)+(0, -1.0)$){$\left( \, \int \, \right)$};

\draw[arrow, red!80!white, dashed] ([xshift=-10mm]J_sub_1.west) -- node[xshift=-0.5mm, above] {$\boldsymbol{x}_{\text{ref},k}$} (J_sub_1);
\draw[arrow, green!70!black, densely dashed] (J_sub_1.east) node[xshift=3.5mm, above] {$\widehat{\boldsymbol{\epsilon}}_k^{\pm}$} -- (gain.west);
\draw[arrow, green!70!black, densely dashed] (J_sum.north) node[xshift=-3mm, yshift=6mm] {$\widehat{\boldsymbol{x}}_k^{\pm}$} -- (J_sub_1.south);
\draw[arrow, green!70!black, densely dashed] (gain.east) node[xshift=4mm, above] {$\boldsymbol{u}_k$} -- (SS_GT.west);
\draw[arrow, green!70!black, densely dashed] ([xshift=3.5mm]gain.east) |- ([yshift=2mm]SS_est.west);
\draw[arrow, blue!45!cyan, dashed] (SS_GT.east) node[xshift=6mm, above] {$\boldsymbol{y}_k$} -- ([xshift=25mm]SS_GT.east);
\draw[arrow, blue!45!cyan, dashed] ([yshift=1.0mm]SS_est.east) node[xshift=6mm, above] {$\widehat{\boldsymbol{y}}_k$} -- (J_sub_2.west);
\draw[arrow, blue!45!cyan, dashed] (J_sub_2.south) node[xshift=-10.5mm, below=11.5mm] {$\widetilde{\boldsymbol{y}}_k$} |- (SS_dif.east);
\draw[arrow, blue!45!cyan, dashed] (SS_dif.west) node[xshift=-27mm, above] {$\boldsymbol{L}_{\widehat{\boldsymbol{x}}_k} \widetilde{\boldsymbol{y}}_k$} -| (J_sum.south);
\draw[arrow, blue!45!cyan, dashed] ([yshift=14.0mm]J_sub_2.north) -- (J_sub_2.north);
\draw[arrow, orange!80!white, densely dashed] ([xshift=-3mm, yshift=-0.5mm]SS_est.west) -- node[xshift=-9.5mm, yshift=3mm] {$\hat{\boldsymbol{x}}^-$} (J_sum.east);
\draw[arrow, orange!80!white, densely dashed] ([yshift=-2mm]SS_est.east) -- ++(1.0,0) node[xshift=-3.5mm, yshift=-5mm] {$\dot{\hat{\boldsymbol{x}}}_k^-$} |- (J_int.east);
\draw[arrow, orange!80!white, densely dashed] (J_int.west) -- ++(-0.95,0) node[xshift=4mm, yshift=2.8mm] {$\hat{\boldsymbol{x}}_k^-$} |- ([yshift=-0.5mm]SS_est.west);

\end{tikzpicture}
\caption{Functional block diagram illustrating the coupled dynamics between the true plant $\mathcal{S}$, estimator $\widehat{\mathcal{S}}$, and error perturbation system $\widetilde{\mathcal{S}}$. The recursive Bayesian update combines prior model propagation (orange) and measurement innovation (blue) at node $\Sigma$ to form the posterior belief estimate (green) driving state-feedback control $\boldsymbol{\kappa}(\cdot)$.}
\label{fig:diagram_S}
\end{figure}

\paragraph{Bifurcated Tracking Error}
Because the true plant state $\boldsymbol{x}_k$ is unobservable, state-feedback controllers must execute using the estimated tracking error
\begin{align} \label{eq:perceived_error}
\hat{\boldsymbol{\epsilon}}_{\boldsymbol{x},k}^{\pm} \triangleq \hat{\boldsymbol{x}}_k^\pm - \boldsymbol{x}_{\mathrm{ref},k} \, .
\end{align}
Fig.~\ref{fig:diagram_S} illustrates this bifurcated estimation mechanism. Under nominal closed-loop operation ($\hat{\boldsymbol{x}}_k^+$), measurement assimilation via $\widetilde{\mathcal{S}}$ bounds posterior uncertainty ($\mathbf{P}_k^+ \le \mathbf{P}_k^-$), preserving NDI cancellation fidelity and error stability. Conversely, during telemetry outages, measurement feedback ceases ($\hat{\boldsymbol{x}}_k^\pm = \hat{\boldsymbol{x}}_k^-$), and prior uncertainty inflates via process noise integration
\begin{align} \label{eq:prior_cov_propagation}
\mathbf{P}_{k+1}^- = \mathbf{F}_{\hat{\boldsymbol{x}}, k} \mathbf{P}_k^- \mathbf{F}_{\hat{\boldsymbol{x}}, k}^\top + \boldsymbol{\Sigma}_{w,k} \, .
\end{align}
This operational duality induces a bifurcated tracking error trajectory $\hat{\boldsymbol{\epsilon}}_{\boldsymbol{x},k}^{\pm}$, directly modulating closed-loop stability across measurement availability regimes.

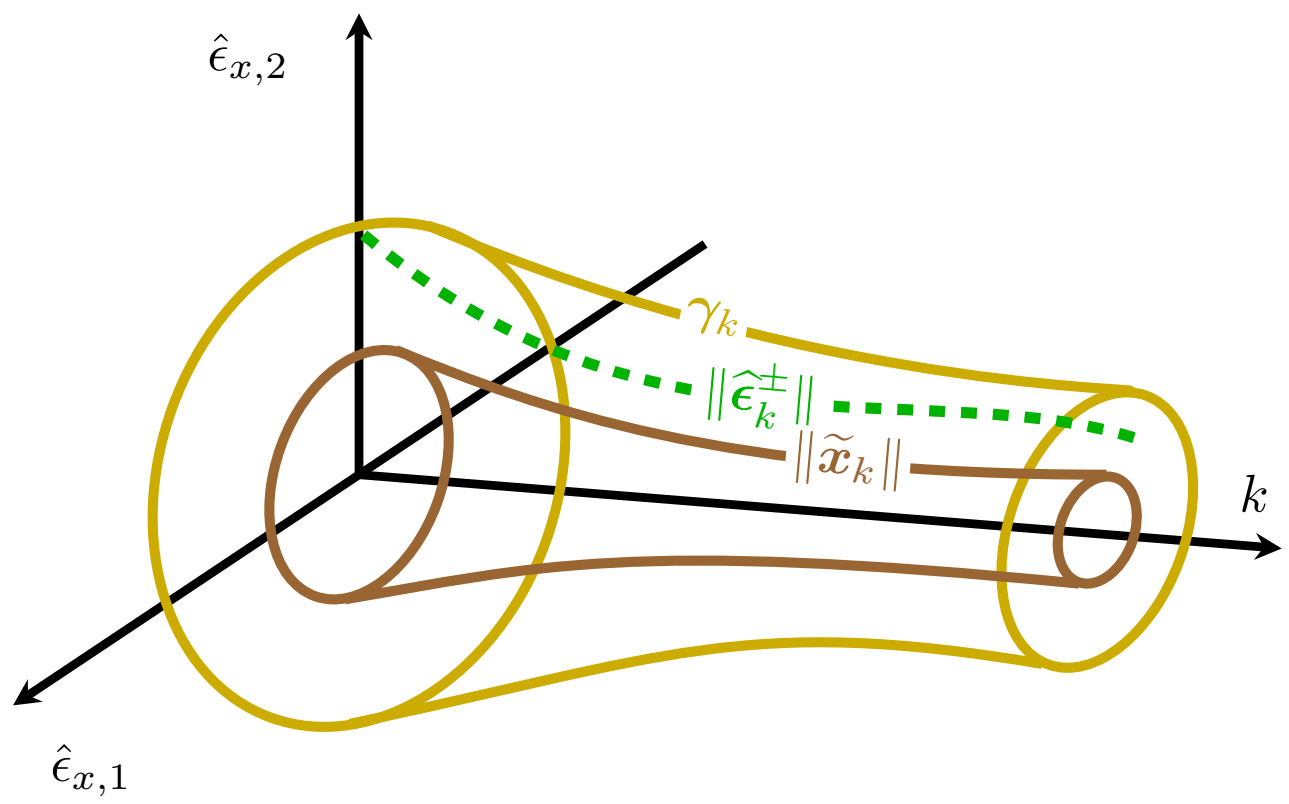
\begin{figure}[b]
\centering
\begin{tikzpicture}[>=Latex,
node distance=16mm and 24mm,
block/.style = {draw, rounded corners=3pt, thick, align=center, inner sep=4pt, minimum width=14mm, minimum height=10mm},
labelnode/.style={font=\normalsize, inner sep=1pt},
every node/.style={font=\normalsize},  
line_ax/.style={-{Stealth[length=5pt, width=5pt]}, line width=1.65pt},
line_in/.style={brown!80!black, line width=1.9pt},
line_ot/.style={gold!80!black, line width=1.9pt},
d_arrow/.style={green!70!black, line width=2.1pt, dashed},
d_line/.style={line width=1.5pt, dashed},
frame/.style={line width=4pt, draw=gray!50},
circ/.style={thick, inner sep=0pt},
scale=\linewidth/0.8\textwidth] 
]
\def\Lx{0}      \def\Ly{0}
\def\Rx{8.0}    \def\Ry{-0.6}
\def\RoX{2.15}  \def\RoY{2.8}
\def\roX{0.9}  \def\roY{1.4}
\def\ROX{0.95}  \def\ROY{1.55}
\def\rOX{0.4}  \def\rOY{0.6}

\coordinate (O)     at (\Lx,\Ly);

\coordinate (LoTop) at (\Lx+\RoX, \Ly+\RoY);
\coordinate (LoBot) at (\Lx+\RoX, \Ly-\RoY);
\coordinate (RoTop) at (\Rx-\ROX, \Ry+\ROY);
\coordinate (RoBot) at (\Rx-\ROX, \Ry-\ROY);

\coordinate (LiTop) at (\Lx+\roX, \Ly+\roY);
\coordinate (LiBot) at (\Lx+\roX, \Ly-\roY);
\coordinate (RiTop) at (\Rx-\rOX, \Ry+\rOY);
\coordinate (RiBot) at (\Rx-\rOX, \Ry-\rOY);

\draw[line_ax] (O) -- ++(0,5);
\draw[line_ax] ($(O)+(3.75,2.5)$) -- ++(-7.5,-5);
\draw[line_ax] (O) -- ++(10.0,-0.8);

\def\RoT{-20}

\draw[line_ot] (\Lx,\Ly) ellipse [x radius=\RoX, y radius=\RoY, rotate=\RoT];
\draw[line_ot] (\Rx,\Ry) ellipse [x radius=\ROX, y radius=\ROY, rotate=\RoT];

\draw[line_ot]($(LoTop)+(-1.4,-0.1)$) .. controls ($(LoTop)+(3,-2)$) and ($(RoTop)+(2.,0.0)$) .. ($(RoTop)+(1.2,-0.05)$);
\draw[line_ot]($(LoBot)+(-2.25,0.1)$) .. controls ($(LoBot)+(1.1,0.8)$) and ($(LoBot)+(2,1.3)$) .. ($(RoBot)+(0.35,0.1)$);

\draw[line_in] (\Lx,\Ly) ellipse [x radius=\roX, y radius=\roY, rotate=\RoT];
\draw[line_in] (\Rx,\Ry) ellipse [x radius=\rOX, y radius=\rOY, rotate=\RoT];

\draw[line_in]($(LiTop)+(-.5,-0.05)$) .. controls ($(LiTop)+(1.1,-0.7)$) and ($(LiTop)+(3,-1.4)$) .. ($(RiTop)+(.5,0)$);
\draw[line_in]($(LiBot)+(-1.05,0.05)$) .. controls ($(LiBot)+(1,+0.4)$) and ($(LiBot)+(2.0,0.7)$) .. ($(RiBot)+(.2,0.02)$);
\draw[d_arrow]($(LoTop)+(-2.1,-0.2)$) .. controls ($(LoTop)+(1,-2.8)$) and ($(LoTop)+(4.0,-1.7)$) .. ($(RiTop)+(0.8,+0.4)$);

\node at ($(O)+(-2.9, -3.2)$) {$ \hat{{\epsilon}}_{x,1} $};
\node at ($(O)+(-1.2, 4.5)$) {$ \hat{{\epsilon}}_{x,2} $};
\node at ($(O)+(9.7, -0.2)$) {$k$};
\node[green!70!black, fill=white, text opacity=1, fill opacity=1.0, rotate=-5, inner sep=2.0pt] at (4.35,0.85) (cs_E){$\| \widehat{\boldsymbol{\epsilon}}_k^{\pm} \| $}; 
\node[brown!80!black, fill=white, text opacity=1, fill opacity=1.0, rotate=-5, inner sep=1.0pt] at (5.3,0.15) (cs_E){$\| \widetilde{\boldsymbol{x}}_k \|$}; 
\node[gold!80!black, fill=white, text opacity=1, fill opacity=1.0, rotate=-10, inner sep=1.0pt] at (3.85,1.7) (cs_E){$\boldsymbol{\gamma}_k$}; 
\end{tikzpicture}
\caption{Evolution of the invariant error tube $\mathcal{T}_{\epsilon, k}$ across discrete steps $k$. On the transverse state plane $(\hat{\epsilon}_{\boldsymbol{x},1}, \hat{\epsilon}_{\boldsymbol{x},2})$, the perceived error trajectory $\|\hat{\boldsymbol{\epsilon}}_{\boldsymbol{x},k}\|$ (green dashed) dissipates within a time-varying annulus. The inner boundary (brown) denotes the estimation error floor $\| \tilde{\boldsymbol{x}}_k \|$, while the outer boundary (dark yellow) denotes the dissipative ceiling $\gamma_k$.}
\label{fig:stab}
\end{figure}

\paragraph{The Estimation Error Floor}
To establish how estimation quality fundamentally limits closed-loop tracking, recall that the true physical tracking error $\boldsymbol{\epsilon}_{\boldsymbol{x},k}$ in \eqref{eq:epsilon_x} relates to the perceived error $\hat{\boldsymbol{\epsilon}}_{\boldsymbol{x},k}$ via the reverse triangle inequality \cite{engelsman2026revisiting}
\begin{align} \label{eq:lower_bound}
\| \boldsymbol{\epsilon}_{\boldsymbol{x},k} \| = \| \hat{\boldsymbol{\epsilon}}_{\boldsymbol{x},k} + \tilde{\boldsymbol{x}}_k \| \geq \Big| \| \hat{\boldsymbol{\epsilon}}_{\boldsymbol{x},k} \| - \|\tilde{\boldsymbol{x}}_k \| \Big| \, .
\end{align}
Consider the positive-definite quadratic Lyapunov candidate
\begin{align} \label{eq:V_Lyp}
V(\hat{\boldsymbol{\epsilon}}_{\boldsymbol{x},k}) = \frac{1}{2}(\hat{\boldsymbol{\epsilon}}_{\boldsymbol{x},k})^\top \mathbf{P}_V \hat{\boldsymbol{\epsilon}}_{\boldsymbol{x},k} \, , 
\end{align}
where $\mathbf{P}_V \succ 0$ solves the discrete Algebraic Lyapunov Equation. The one-step Lyapunov difference along closed-loop trajectories satisfies \cite{lyapunov_1, Jing2020}
\begin{align} \label{eq:V_dot_neg}
\Delta V_k \le \underbrace{ -\lambda \|\hat{\boldsymbol{\epsilon}}_{\boldsymbol{x},k}\|^2 }_{\text{Dissipation}} + \underbrace{ \mu \|\tilde{\boldsymbol{x}}_k\|^2 + d_k }_{\text{Excitation}} \, ,
\end{align}
where $\lambda > 0$ is the discrete dissipation rate, $\mu > 0$ is the controller-estimator coupling factor, and $d_k \ge 0$ bounds external disturbances. Under bounded estimation error $\|\tilde{\boldsymbol{x}}_k\|$, the perceived tracking error remains uniformly ultimately bounded (UUB), converging to an invariant set with dissipative ceiling \cite{Barmish1983}
\begin{align} \label{eq:upper_bound}
\gamma_k \triangleq \sqrt{ \frac{ \mu \| \tilde{\boldsymbol{x}}_k \|^2 + d_k }{\lambda} } \, .
\end{align}
\begin{corollary} \label{cor:tube}
Reconciling the uncertainty lower bound \eqref{eq:lower_bound} with the dissipative ceiling \eqref{eq:upper_bound} yields the forward-invariant confinement tube $\mathcal{T}_{\epsilon,k}$ for the true physical tracking error
\begin{align} \label{eq:Inv_tube} 
\mathcal{T}_{\epsilon,k} \triangleq \left\{ \, \hat{\boldsymbol{\epsilon}}_k \in \mathbb{R}^n : \| \tilde{\boldsymbol{x}}_k \| \le \| \hat{\boldsymbol{\epsilon}} \| \le \gamma_k \, \right\} \, .
\end{align}
\end{corollary}
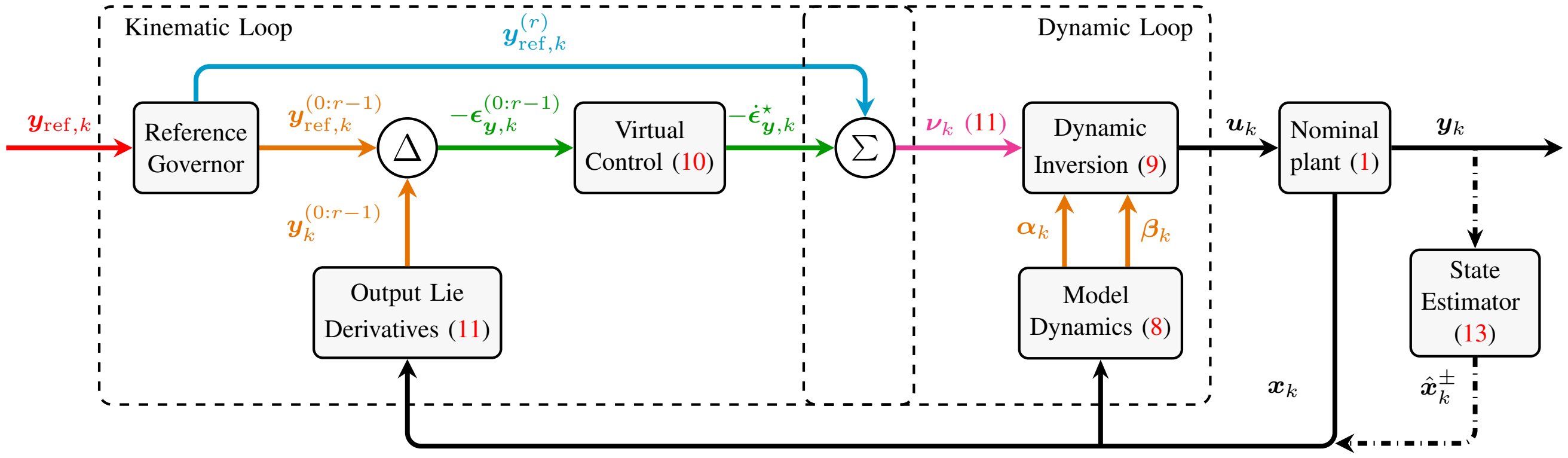
\begin{figure*}[t]
\centering
\begin{tikzpicture}[ 
  block/.style={draw, thick, minimum width=12mm, minimum height=1cm, fill=gray!6, rounded corners=1mm},
  block_small/.style={draw, thick, minimum width=12mm, minimum height=7mm, fill=gray!6, rounded corners=1mm},
  joint/.style={draw, circle, thick, minimum size=6.5mm, inner sep=0pt},
  arrow/.style={-{Stealth[length=6pt, width=6pt]}, rounded corners=1.5mm, line width=1.5pt},
  labelnode/.style={font=\footnotesize, inner sep=1pt},
  every node/.style={font=\footnotesize}
]

\node[block] (mod_ref) {\shortstack{Reference\vspace{.5mm}\\Governor}};
\node[joint, right=13mm of mod_ref] (J_sub_1) {\large$\Delta$};
\node[block, right=15mm of J_sub_1] (mod_gain) {\shortstack{Virtual\vspace{.5mm}\\Control \eqref{eq:error_diff_eq}}};
\node[joint, right=12mm of mod_gain] (J_sum_1) {$\sum$};
\node[block, right=14mm of J_sum_1] (mod_di) {\shortstack{Dynamic\vspace{.5mm}\\Inversion \eqref{eq:control_law_NDI}}};
\node[block, right=11mm of mod_di] (plant) {\shortstack{Nominal\vspace{.5mm}\\plant \eqref{eq:state_space_GT}}};
\node[block, below right=6mm and 2mm of plant] (estimator) {\shortstack{State\vspace{.5mm}\\Estimator \vspace{.5mm}\\ \eqref{eq:state_space_est}}};
\node[block, below=8mm of mod_di] (non_cancel) {\shortstack{Model \vspace{.5mm}\\ Dynamics \eqref{eq:normal_form_discrete}}};
\node[block, below=9.7mm of J_sub_1] (Lie_alpha) {\shortstack{Output Lie\vspace{.5mm}\\ Derivatives \eqref{eq:decoupled_integrator}}};

\draw[arrow, red] ([xshift=-14mm]mod_ref.west) -- node[xshift=-1mm, above] {$\boldsymbol{y}_{\mathrm{ref},k}$} (mod_ref);
\draw[arrow, orange!90!black] (mod_ref) -- node[xshift=2mm, above] {$\boldsymbol{y}_{\mathrm{ref},k}^{(0:r-1)}$} (J_sub_1);
\draw[arrow, green!60!black] (J_sub_1) -- node[xshift=0mm, above] {$- \boldsymbol{\epsilon}_{ \boldsymbol{y},k}^{(0:r-1)}$} (mod_gain);
\draw[arrow, green!60!black] (mod_gain) -- node[xshift=-2mm, above] {$- \dot{\boldsymbol{\epsilon}}_{\boldsymbol{y},k}^\star$} (J_sum_1);
\draw[arrow, cyan!80!black](mod_ref) -- ++(0,0.9) -| node[xshift=-36.5mm, above]{$\boldsymbol{y}_{\mathrm{ref},k}^{(r)}$}(J_sum_1);
\draw[arrow, magenta!90!white] (J_sum_1) -- node[xshift=0.8mm, above] {$\boldsymbol{\nu}_k$ \eqref{eq:decoupled_integrator}} (mod_di);
\draw[arrow] (mod_di) -- node[xshift=1.5mm, above] {$\boldsymbol{u}_k$} (plant);

\draw[arrow] (plant.east) -- ++(19mm,0) node[pos=0.35, above] {$\boldsymbol{y}_k$};
\draw[arrow](plant) -- ++(0,-33mm) -| node[xshift=97mm, yshift=6.3mm]{$\boldsymbol{x}_k$}(Lie_alpha.south);
\draw[arrow, orange!90!black] (Lie_alpha.north) -- node[xshift=-8mm] {$\boldsymbol{y}^{(0:r-1)}_k$} (J_sub_1.south);
\draw[arrow, orange!90!black] ([xshift=-4mm]non_cancel.north) -- node[left] {$\boldsymbol{\alpha}_k$} ([xshift=-4mm]mod_di.south);
\draw[arrow, orange!90!black] ([xshift=3mm]non_cancel.north) -- node[right] {$\boldsymbol{\beta}_k$} ([xshift=3mm]mod_di.south);
\draw[dashdotted, <-, >={Stealth[length=6pt, width=5pt]}, line width=1.5pt] (estimator.north) -- ++(0,11mm) node[above] {};
\draw[dashdotted, arrow] (estimator.south) -- ++(0,-9.5mm) -- ++(-15.5mm,0) node[xshift=11.5mm, yshift=6.2mm] {$\hat{\boldsymbol{x}}_k^{\pm}$};
\draw[<-, >={Stealth[length=6pt, width=5pt]}, line width=1.5pt] (non_cancel.south) -- ++(0,-9.5mm) node[above] {};
%
\draw[dashed, thick, rounded corners, text opacity=1]($(mod_ref.north west)+(-0.37,1.05)$) rectangle node[xshift=-3.3cm, yshift=1.95cm]{Kinematic Loop}  ($(J_sum_1.south east)+(0.3,-2.6)$);
\draw[dashed, thick, rounded corners, text opacity=1]($(J_sum_1.north west)+(-0.44,1.325)$) rectangle node[xshift=1.2cm, yshift=1.95cm]{Dynamic Loop}  ($(non_cancel.south east)+(0.3,-0.498)$);
\end{tikzpicture}
\caption{Failure of implicit determinism under estimator uncertainty: Solid paths show nominal feedback under ideal certainty equivalence (${\boldsymbol{x}}_k$), while dashed paths expose unmodeled loop closure via state estimates ($\hat{\boldsymbol{x}}_k^{\pm}$). Color coding highlights structural feedback paths: reference path (red), feedforward signal (cyan), kinematic output predictions (orange), tracking error compensation (green), virtual control synthesis (magenta), and plant dynamics inversion (black).}
\label{fig:system}
\end{figure*}
\begin{table*}[b]
\caption{Overview of nonlinear control paradigms, underlying mechanisms, and their EA extensions evaluated in this work.}
\label{t:paradigms}
\centering
\footnotesize
\renewcommand{\arraystretch}{1.6}
\begin{tabular}{|m{3.2cm}|m{5.0cm}|c|c|}
\hline
\textbf{Control Approach} \CC & \textbf{Functional Logic} \CC & \textbf{Baseline Control Law ($\boldsymbol{u}_k$)} \CC & \textbf{EA Extension ($\boldsymbol{u}_{\textit{EA},k}$)} \CC \\ \hline
{Feedback Linearization (FL)} \cite{voos2009nonlinear,das2009dynamic,freddi2011feedback} & 
Algebraic input–output map inversion to decouple system couplings and enforce linear error dynamics. & 
$\begin{aligned} \boldsymbol{u}_k &= \boldsymbol{\alpha}_k + \boldsymbol{\beta}_k \boldsymbol{\nu}_k \\ \boldsymbol{\alpha}_k &\triangleq -\mathbf{g}^{\dagger} \boldsymbol{f}(\boldsymbol{x}_k) \\ \boldsymbol{\beta}_k &\triangleq \mathbf{g}^{\dagger} \end{aligned}$ & 
$\begin{aligned} \boldsymbol{u}_{\textit{EA},k} &= \hat{\boldsymbol{\alpha}}_k + \hat{\boldsymbol{\beta}}_k \boldsymbol{\nu}_k + \boldsymbol{\Delta}(\boldsymbol{\eta}_k) \\ \hat{\boldsymbol{\alpha}}_k &\triangleq \boldsymbol{\alpha}(\hat{\boldsymbol{x}}_k) \\ \hat{\boldsymbol{\beta}}_k &\triangleq \boldsymbol{\beta}(\hat{\boldsymbol{x}}_k) \end{aligned}$ \\ \hline
{Geometric Control (GC)} \cite{lee2010geometric,sreenath2013geometric,goodarzi2015geometric} & 
Coordinate-free tracking synthesis, using intrinsic position and attitude errors to define globally consistent control actions. & 
$\boldsymbol{u}_k = -\mathbf{K} \begin{pmatrix} \boldsymbol{\epsilon}_{\text{pos},k} \\ \boldsymbol{\epsilon}_{\text{att},k} \end{pmatrix} + \boldsymbol{u}_{\text{ff},k}$ & 
$\boldsymbol{u}_{\textit{EA},k} = -\mathbf{K}(\boldsymbol{\eta}_k) \begin{pmatrix} \hat{\boldsymbol{\epsilon}}_{\text{pos},k} \\ \hat{\boldsymbol{\epsilon}}_{\text{att},k} \end{pmatrix} + \boldsymbol{u}_{\text{ff},k}$ \\ \hline
{Cascaded Control (CC)} \cite{zhang2017cascaded,smeur2018cascaded,xu2020backstepping,CABUK2026111844} & 
Hierarchical time-scale separation into an outer kinematic loop (slow) and an inner dynamic control loop (fast). & 
$\begin{aligned} \boldsymbol{u}_k &= \boldsymbol{\pi}_{\text{in}}(\boldsymbol{\epsilon}_{\text{att},k}, \boldsymbol{\nu}_k) \\ \boldsymbol{\nu}_k &= \boldsymbol{\pi}_{\text{out}}(\boldsymbol{\epsilon}_{\text{pos},k}) \end{aligned}$ & 
$\begin{aligned} \boldsymbol{u}_{\textit{EA},k} &= \boldsymbol{\pi}_{\text{in}}(\hat{\boldsymbol{\epsilon}}_{\text{att},k}, \boldsymbol{\nu}_k, \boldsymbol{\eta}_k) \\ \boldsymbol{\nu}_k &= \boldsymbol{\pi}_{\text{out}}(\hat{\boldsymbol{\epsilon}}_{\text{pos},k}, \boldsymbol{\eta}_k) \end{aligned}$ \\ \hline
{Differential Flatness (DF)} \cite{faessler2017differential,morrell2018differential,tal2020accurate} & 
Parameterization of system states and inputs via a flat output and its derivatives, enabling direct trajectory generation. & 
$\begin{aligned} \boldsymbol{x}_k &= \Phi(\boldsymbol{y}_k^{[0:r]}) \\ \boldsymbol{u}_k &= \Psi(\boldsymbol{y}_k^{[0:r]}) \end{aligned}$ & 
$\begin{aligned} \hat{\boldsymbol{x}}_k &= \Phi(\hat{\boldsymbol{y}}_k^{[0:r]}) \\ \boldsymbol{u}_{\textit{EA},k} &= \Psi(\hat{\boldsymbol{y}}_k^{[0:r]}) + \boldsymbol{\Delta}(\boldsymbol{\eta}_k) \end{aligned}$ \\ \hline
{Nonlinear MPC (NMPC)} \cite{grune2016nonlinear, kang2009linear,wang2021efficient,LIU2026111657} & 
Receding-horizon constrained optimization subject to discrete nonlinear dynamics and bounded state-input sets. & 
$\begin{aligned} \boldsymbol{u}_k^* = \arg \min_{\boldsymbol{u} \in \mathcal{U}} \sum_{j=0}^{N-1} \ell_c (\cdot) \\ \text{s.t.} \ \boldsymbol{x}_{k+j+1} = \boldsymbol{f}(\boldsymbol{x}_{k+j}, \boldsymbol{u}_{k+j}) \end{aligned}$ & 
$\begin{aligned} \boldsymbol{u}_{\textit{EA},k}^* = \arg \min_{\boldsymbol{u} \in \mathcal{U}} \sum_{j=0}^{N-1} \ell_c (\cdot , \boldsymbol{\eta}_k) \\ \text{s.t.} \ \hat{\boldsymbol{x}}_{k+j+1} = \boldsymbol{f}(\hat{\boldsymbol{x}}_{k+j}, \boldsymbol{u}_{k+j}) \end{aligned}$ \\ \hline
{Incremental NDI (INDI)} \cite{smeur2016adaptive,di2016modeling,STEINERT2025103553} & 
Incremental actuation based on measured state acceleration increments ($\Delta \boldsymbol{\nu}_k$) to reduce model dependency. & 
$\begin{aligned} \boldsymbol{u}_k &= \boldsymbol{u}_{k-1} + \mathbf{g}^{\dagger} \Delta \boldsymbol{\nu}_k \\ \Delta \boldsymbol{\nu}_k &\triangleq \boldsymbol{\nu}_k - \boldsymbol{\delta x}_k \end{aligned}$ & 
$\begin{aligned} \boldsymbol{u}_{\textit{EA},k} &= \boldsymbol{u}_{k-1} + \hat{\mathbf{g}}^{\dagger} \Delta \hat{\boldsymbol{\nu}}_k \\ \Delta \hat{\boldsymbol{\nu}}_k &\triangleq \boldsymbol{\nu}_k - \boldsymbol{\delta}\hat{\boldsymbol{x}}_k + \boldsymbol{\Delta}(\boldsymbol{\eta}_k) \end{aligned}$ \\ \hline
\end{tabular}
\end{table*}
As illustrated in Fig.~\ref{fig:stab}, this geometry captures the dynamic tension between control dissipation and perception uncertainty. While the feedback policy contracts error trajectories toward the outer ceiling $\gamma_k$, the estimation residue $\|\tilde{\boldsymbol{x}}_k\|$ bounds this contraction from within, forming the forward-invariant tube $\mathcal{T}_{\epsilon,k}$.

\begin{remark}[Uncompensable Tracking Floor]
\label{rem:perception_floor}
The inner radius of the invariant error tube $\mathcal{T}_{\epsilon,k}$ is strictly dictated by the state estimation error norm $\|\tilde{\boldsymbol{x}}_k^\pm\|$. Consequently, feedback control gains $\boldsymbol{\kappa}(\cdot)$ can attenuate the dissipative ceiling $\gamma_k$, but cannot drive the physical tracking error below this Bayesian uncertainty floor ($\|\boldsymbol{\epsilon}_{\boldsymbol{x},k}\| \to 0$).
\end{remark}

Fig.~\ref{fig:system} depicts the baseline tracking architecture in modern feedback linearization and dynamic inversion approaches. While ideal formulations assume state determinism via exact full-state feedback (solid paths), practical implementations rely on estimated state feedback ($\hat{\boldsymbol{x}}_k^{\pm}$, dashed paths). The system alternates between propagation ($\hat{\boldsymbol{x}}_k^-$) during sensor outages and correction ($\hat{\boldsymbol{x}}_k^+$) upon measurement availability. This dynamic coupling feeds estimator transients directly into the control loop, breaking the nominal decoupling of tracking errors \cite{Molin2013}.

\section{Proposed Methodology} \label{sec:method}
Grounded in the structural bounds of Remark~\ref{rem:perception_floor}, we now synthesize the \emph{estimation-aware} control operator. The proposed formulation is inherently paradigm-agnostic; it augments any baseline nonlinear control policy by directly embedding posterior estimation uncertainty into the feedback loop. 

\subsection{Architectural Integration of Estimation Awareness} \label{subsec:family}
To preserve anisotropic sensitivity along weakly observed state directions, we define the generalized uncertainty metric $\boldsymbol{\eta}_k \triangleq \boldsymbol{\Delta}(\mathbf{P}_{k\mid k})$, where $\boldsymbol{\Delta}(\cdot)$ maps the posterior error covariance $\mathbf{P}_{k\mid k}$ onto the control manifold. 
\\
Table~\ref{t:paradigms} surveys representative nonlinear control paradigms ($\boldsymbol{u}_k$) alongside their corresponding EA extensions ($\boldsymbol{u}_{\textit{EA},k}$) tailored for high-bandwidth aerial systems \cite{li2015survey, engelsman2025guidance, STEINERT2025103553}. These formulations are categorized across three functional layers:
\begin{enumerate}[label=\roman*), itemsep=1.5pt, parsep=1pt, topsep=2pt]
\item \textbf{Signal-Level Integration:}
This layer injects a dynamic robustifying signal $\boldsymbol{\Delta}_{\mathrm{sig}}(\boldsymbol{\eta}_k)$ directly into the virtual or physical control command. By explicitly compensating for unmodeled residual dynamics induced by state estimation errors ($\hat{\boldsymbol{x}}_k \neq \boldsymbol{x}_k$), $\boldsymbol{\Delta}_{\mathrm{sig}}(\boldsymbol{\eta}_k)$ acts as an active feedforward term that mitigates divergence between perceived and true physical plant dynamics.

\item \textbf{Sensitivity-Level Integration:}
Here, $\boldsymbol{\eta}_k$ actively modulates controller bandwidth via an uncertainty-dependent gain operator $\mathbf{K}(\boldsymbol{\eta}_k)$. This mechanism regulates closed-loop sensitivity by automatically throttling feedback authority along weakly observed state directions when estimation confidence degrades, preventing noise amplification and parasitic chatter in the actuation channels.

\item \textbf{Topological-Level Integration:}
In optimization-based paradigms (e.g., NMPC), awareness is embedded directly into the state-space geometry. The metric $\boldsymbol{\eta}_k$ dictates the admissible boundary of the safe operating region by dynamically contracting the state constraint set $\mathcal{X}(\boldsymbol{\eta}_k)$ or regularizing the stage cost $\ell_c(\cdot)$. By updating the error manifold under estimator variance, this layer preserves robust forward invariance for the true physical state $\boldsymbol{x}_k$.
\end{enumerate}

\subsection{Estimation-Aware Mechanization}
As highlighted in the bottom row of Table~\ref{t:paradigms}, the proposed EA strategy embeds the augmentation operator $\boldsymbol{\Delta}(\boldsymbol{\eta}_k)$ directly into the nominal INDI control law:
\begin{align} \label{eq:u_EA}
\begin{aligned} 
\boldsymbol{u}_{\textit{EA},k} &= \boldsymbol{u}_{k-1} + \hat{\mathbf{g}}^{\dagger} \Delta \hat{\boldsymbol{\nu}}_k \, , \\ 
\Delta \hat{\boldsymbol{\nu}}_k &\triangleq \boldsymbol{\nu}_k - \boldsymbol{\delta}\hat{\boldsymbol{x}}_k + \boldsymbol{\Delta}(\boldsymbol{\eta}_k) \, . 
\end{aligned}
\end{align}
This estimation awareness is embedded by transforming the error covariance matrix $\mathbf{P}_k$ into an anisotropic gating operator that selectively scales and cross-projects estimates via
\begin{align} \label{eq:awareness}
\begin{aligned}
\boldsymbol{\Delta}(\boldsymbol{\eta}_k) &= \left( \mathbf{I} - \overline{\mathbf{P}}_k \right) \boldsymbol{\delta}\hat{\boldsymbol{x}}_k \, ,  \\
\overline{\mathbf{P}}_k &\triangleq \boldsymbol{D}_k^{-1/2} \mathbf{P}_k \boldsymbol{D}_k^{-1/2} \, , 
\end{aligned}
\end{align}
where $\boldsymbol{D}_k = \mathrm{diag}(\mathbf{P}_k)$ and $\overline{\mathbf{P}}_k$ denotes the normalized correlation matrix. In operational regimes characterized by high estimation certainty, off-diagonal state correlations remain minimal ($\overline{\mathbf{P}}_k \to \mathbf{I}$), driving $\boldsymbol{\Delta}(\boldsymbol{\eta}_k) \to \mathbf{0}$ and seamlessly recovering the nominal INDI control law. Conversely, during high-uncertainty transients, off-diagonal coupling terms in $(\mathbf{I} - \overline{\mathbf{P}}_k)$ expand, cross-projecting state difference increments $\boldsymbol{\delta}\hat{\boldsymbol{x}}_k$ to attenuate $\Delta \hat{\boldsymbol{\nu}}_k$ and prevent aggressive actuation on unobservable or corrupted state estimates.

\subsection{Nonlinear System Dynamics} \label{sec:system}
Focusing on quadrotor flight dynamics, we establish the continuous-time open-loop equations of motion governing the 6-DoF platform. To maintain consistency with modern flight control literature \cite{hoffmann2007quadrotor, engelsman2026czupt, yu2020two}, we adopt a standard control-affine state-space formulation.

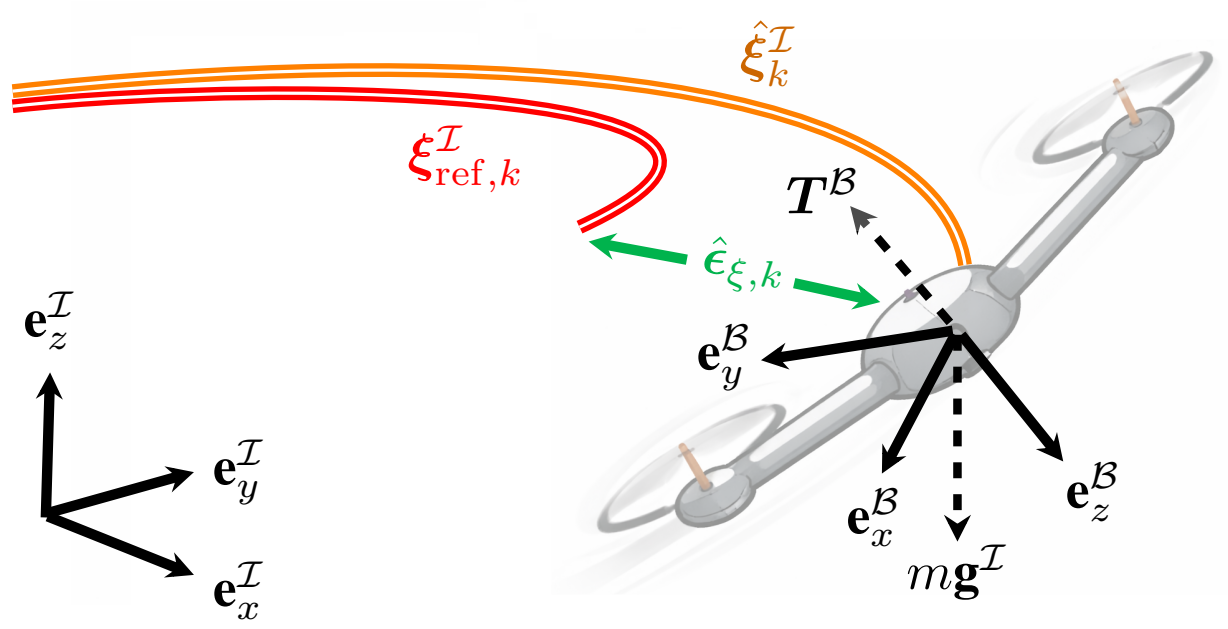
\begin{figure}[!h]
\centering
\begin{tikzpicture}[>=Latex,
node distance=16mm and 24mm,
block/.style = {draw, rounded corners=3pt, thick, align=center, inner sep=4pt, minimum width=14mm, minimum height=10mm},
labelnode/.style={font=\normalsize, inner sep=1pt},
every node/.style={font=\normalsize},  
arrow/.style={-{Stealth[length=5pt, width=5pt]}, line width=1.95pt},
d_arrow/.style={-{Stealth[length=5pt, width=5pt]}, line width=1.6pt, dashed},
d_line/.style={line width=1.5pt, dashed},
frame/.style={line width=4pt, draw=gray!50},
circ/.style={thick, inner sep=0pt},
]
\node[inner sep=0pt, align=center, opacity=0.5] (quad) {\includegraphics[width=0.485\textwidth]{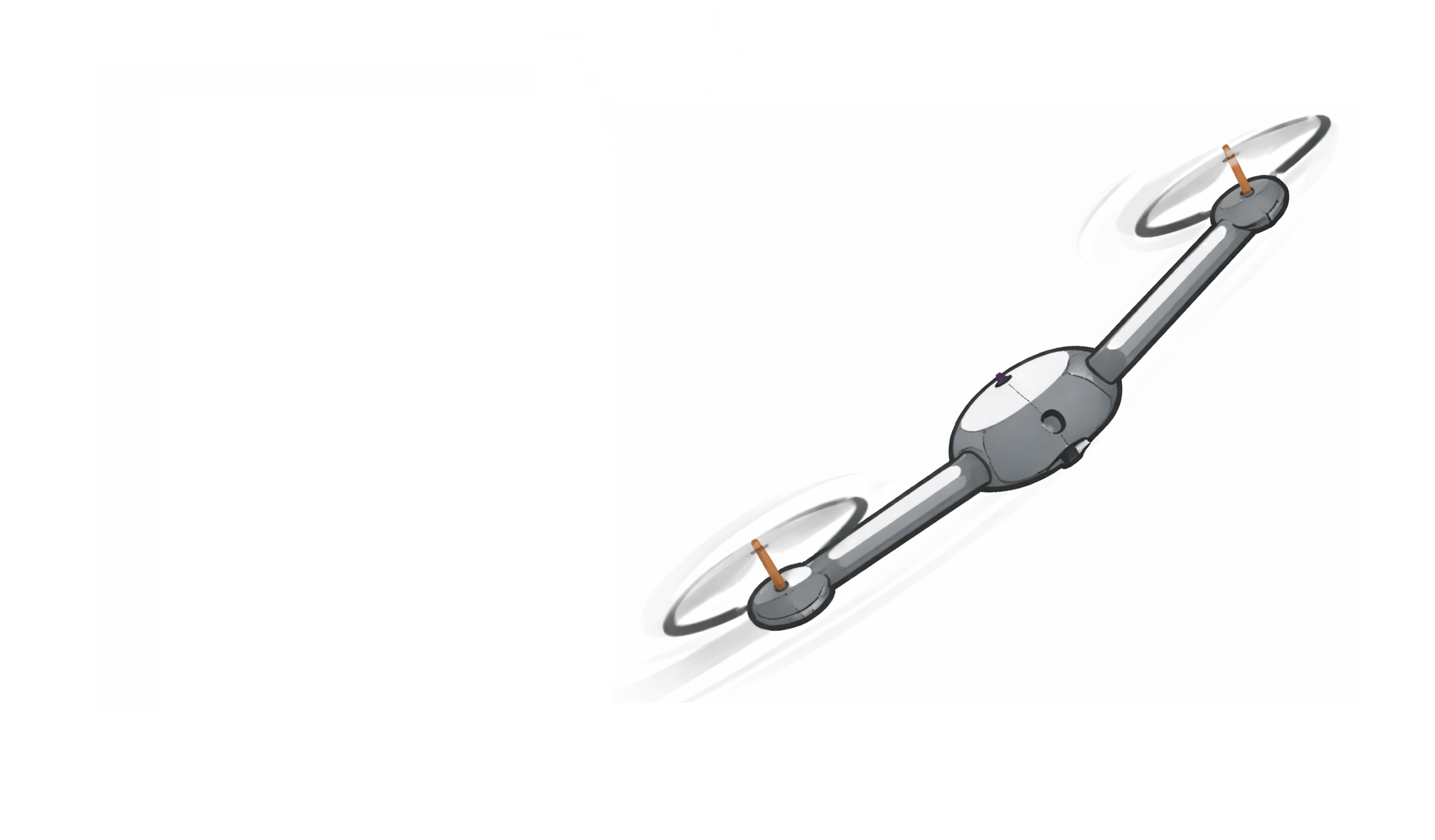}};

\node[circ, shift={(+20mm,+15mm)}] at (-6,-2.7) (cs_I){};
\draw[arrow] (cs_I) --++ (1.25*.8,-.5*.8) node[shift={(3mm,-1mm)}]{$\textbf{e}_x^\mathcal{I}$};
\draw[arrow] (cs_I) --++ (1.25*.8,.35*.8) node[shift={(3mm,0mm)}]{$\textbf{e}_y^\mathcal{I}$};
\draw[arrow] ($(cs_I)+(0.02,-0.02)$) --++ (0.035*.8, 1.2*.8) node[above]{$\textbf{e}_z^\mathcal{I}$};

\node[circ, shift={(+0mm,-2.5mm)}] at (2.05,0.27) (cs_B){};
\draw[d_arrow] ($(cs_B)+(-.05,0.05)$) --++ (-0.66,0.77) node[shift={(-1.8mm,1.0mm)}, fill=white, text opacity=1, fill opacity=0.3]{${\boldsymbol{T}^\mathcal{B}}$};
\draw[arrow] (cs_B) --++ (-1.3,-0.2) node[shift={(-2.4mm,0.2mm)}]{$\textbf{e}_y^\mathcal{B}$};
\draw[arrow] (cs_B) --++ (-0.5,-0.95) node[shift={(-0.5mm,-2.9mm)}]{$\textbf{e}_x^\mathcal{B}$};
\draw[arrow] (cs_B) --++ (+.7,-.87) node[shift={(2mm,-2.mm)}]{$\textbf{e}_z^\mathcal{B}$};
\draw[d_arrow] (cs_B) --++ (0,-1.4) node[shift={(0mm,-2mm)}]{$m \textbf{{g}}^\mathcal{I}$};
%
\node[circ, blue!30!green, shift={(+2mm,+1mm)}, rotate=-5] at (.44,0.35) (cs_E){$\ \hat{\boldsymbol{\epsilon}}_{\xi,k} \ $}; 
\draw[arrow, blue!30!green]($(cs_E)+(0.33,-0.12)$) --++ (0.57, -0.125) node[above]{};
\draw[arrow, blue!30!green]($(cs_E)+(-0.38,0.035)$) --++ (-0.65, .12) node[above]{};
\node at ($(cs_I)+(0,-.5)$) [orange!80!black, fill=white, text opacity=1, fill opacity=0.0, shift={(48mm,35.5mm)}]{$\hat{\boldsymbol{\xi}}^\mathcal{I}_k$};
\node at ($(cs_I)+(0,-.5)$) [red, fill=white, text opacity=1, fill opacity=0.0, shift={(28mm,28.5mm)}]{$\boldsymbol{\xi}^\mathcal{I}_{\mathrm{ref},k} $};

\draw[orange, line width=1.0pt, double]($(-4.2,1.6)$) .. controls ($(-1.5,1.9)$) and  ($(1.9,1.8)$) .. ($(2.1,0.45)$);
\draw[red, line width=1.0pt, double]($(-4.2,1.5)$) .. controls ($(-2.4,1.7)$) and  ($(1.5,1.6)$) .. ($(-0.45,0.7)$);
\end{tikzpicture}
\caption{Free-body diagram of the 6-DoF quadrotor platform, illustrating the reference trajectory (red), estimated path (green), and corresponding tracking error vector (green).} 
\label{fig:quad_manu}
\end{figure}

As illustrated in Fig.~\ref{fig:quad_manu}, the vehicle is modeled as a 6-DoF rigid body with mass $m$ and moment of inertia matrix $\boldsymbol{J}$. To eliminate representation singularities, its orientation is parameterized by the unit quaternion $\boldsymbol{q} \in \mathbb{S}^3$, where $\boldsymbol{R}(\boldsymbol{q}) \in SO(3)$ denotes the body-to-inertial rotation matrix. Defining $\boldsymbol{\xi}^\mathcal{I} \in \mathbb{R}^3$ as inertial position, $\boldsymbol{v}^\mathcal{B} \in \mathbb{R}^3$ as body linear velocity, and $\boldsymbol{\omega}^\mathcal{B} \in \mathbb{R}^3$ as body angular velocity, the full state vector is expressed as
\begin{align} \label{eq:states}
\boldsymbol{x} \triangleq \begin{bmatrix}
(\boldsymbol{\xi}^\mathcal{I})^\top &
(\boldsymbol{v}^\mathcal{B})^\top &
\boldsymbol{q}^\top &
(\boldsymbol{\omega}^\mathcal{B})^\top
\end{bmatrix}^\top \in \mathbb{R}^{13} \, .
\end{align}
The control input vector $\boldsymbol{u} \triangleq [u_z^\mathcal{B}, (\boldsymbol{u}_\tau^\mathcal{B})^\top]^\top \in \mathbb{R}^4$ collects normalized vertical thrust $u_z^\mathcal{B}$ and body moment commands $\boldsymbol{u}_\tau^\mathcal{B} = [\tau_x^\mathcal{B}, \tau_y^\mathcal{B}, \tau_z^\mathcal{B}]^\top$. 
Combining quadrotor kinematics and rigid-body dynamics under external aerodynamic forces $\boldsymbol{f}_{\mathrm{aero}}^\mathcal{B}$ and moments $\boldsymbol{\tau}_{\mathrm{aero}}^\mathcal{B}$ yields the physical plant in canonical control-affine form \eqref{eq:state_space_GT} as
\begin{align} \label{eq:x_dot_big}
\dot{\boldsymbol{x}} = \begin{pmatrix}
\boldsymbol{R}(\boldsymbol{q}) \boldsymbol{v}^\mathcal{B} \\
\boldsymbol{R}^\top(\boldsymbol{q}) \mathbf{g}^\mathcal{I} - \boldsymbol{\omega}^\mathcal{B} \times \boldsymbol{v}^\mathcal{B} + \frac{1}{m} \left( u_z^\mathcal{B} \mathbf{e}_z + \boldsymbol{f}_{\mathrm{aero}}^\mathcal{B} \right) \\
\frac{1}{2}\boldsymbol{q} \odot \begin{pmatrix} 0 \\ \boldsymbol{\omega}^\mathcal{B} \end{pmatrix} \\
\boldsymbol{J}^{-1} \left( -\boldsymbol{\omega}^\mathcal{B} \times \boldsymbol{J}\boldsymbol{\omega}^\mathcal{B} + \boldsymbol{u}_\tau^\mathcal{B} + \boldsymbol{\tau}_{\mathrm{aero}}^\mathcal{B} \right)
\end{pmatrix} ,
\end{align}
where $\odot$ denotes Hamilton quaternion multiplication and $\mathbf{g}^\mathcal{I} = [0, 0, -\text{g}]^\top$ is gravity vector. 
While the physical plant evolves in continuous time $t$, control synthesis and state estimation are executed at discrete sampling steps $k$.

\section{Analysis and Results} \label{sec:results}
This section assesses the tracking performance of the proposed EA framework under severe model uncertainty and state estimation degradation. To provide clear physical intuition, Section~\ref{subsec:A} first analyzes the nominal trajectory and state evolution of a single baseline INDI controller, while Section~\ref{subsec:B} evaluates the corresponding real-time filter convergence. Then, Section~\ref{subsec:C} analyzes the closed-loop uncertainty mechanics of the EA operator mid-flight. Lastly, Section~\ref{subsec:D} benchmarks all six control paradigms to demonstrate the performance gains and generality of the proposed framework.
\begin{figure}[!t]
\begin{center}
\includegraphics[width=0.46\textwidth]{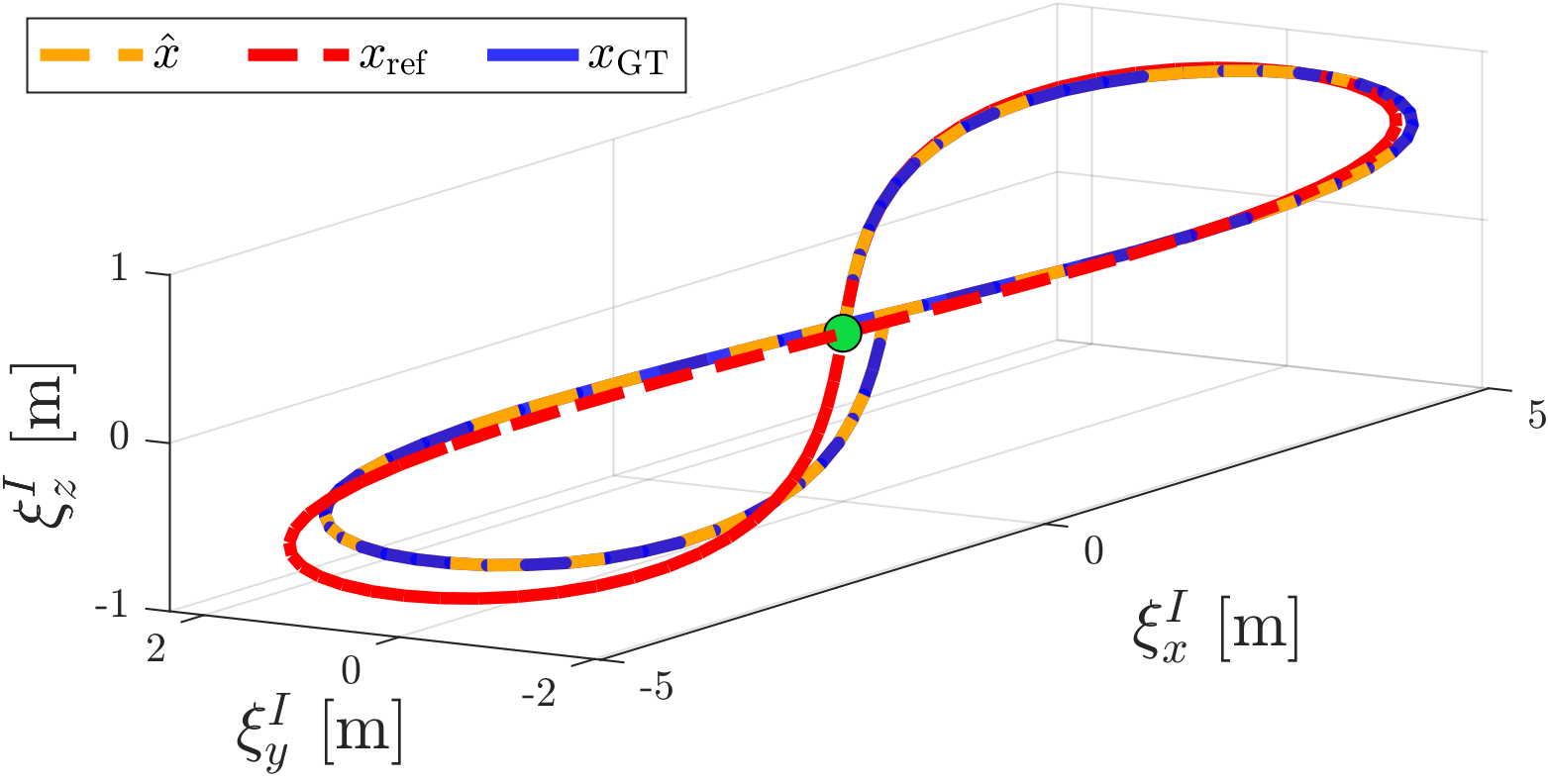}
\caption{Spatial evolution of the reference, true, and estimated trajectories, under nominal tracking at $\mathrm{v}_{\mathrm{ref}} = 1.0\,\text{m/s}$.}
\label{fig:lemni_sub}
\end{center}
\end{figure} 
\begin{figure}[!b]
\begin{center}
\includegraphics[width=0.49\textwidth]{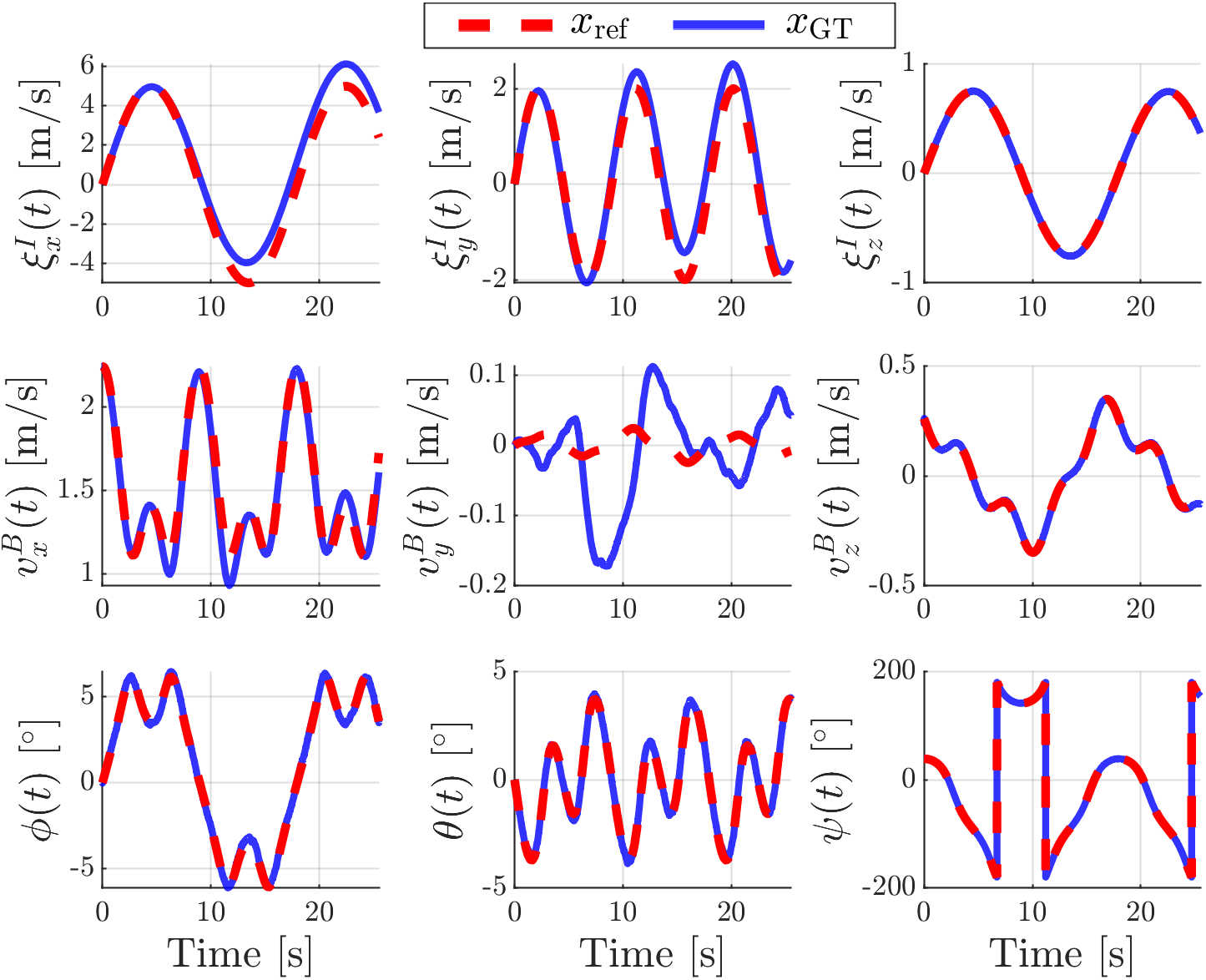}
\caption{States evolution (blue) vs. reference trajectory (red).}
\label{fig:states_1}
\end{center}
\end{figure} 

\subsection{Nominal Baseline Scenario} \label{subsec:A}
The numerical evaluation utilizes a trajectory designed to excite all dynamic modes, under nominal INDI control. While standard circular trajectories exhibit constant curvature with smooth higher-order derivatives (i.e., jerk $\dddot{\boldsymbol{\xi}}$ and snap $\ddddot{\boldsymbol{\xi}}$), we select a 3D tilted lemniscate trajectory. Bounded within a $2\mathcal{R} \times \mathcal{R} \times 0.5 \mathcal{R}$ volume ($\mathcal{R} = 5\,\text{m}$) and inclined at $30^\circ$, this complex spatial path continuously varies curvature and introduces strong cross-axis coupling, thoroughly challenging the underactuated system dynamics.
\\
Fig.~\ref{fig:lemni_sub} illustrates the spatial trajectories, comparing the reference path ($\boldsymbol{\xi}^{\mathcal{I}}_{\mathrm{ref}}$, dashed red), state ground truth ($\boldsymbol{\xi}^{\mathcal{I}}$, blue), and state estimate ($\hat{\boldsymbol{\xi}}^{\mathcal{I}}$, orange). Under nominal operating conditions at a moderate translational velocity ($\mathrm{v}_{\mathrm{ref}} = 1.0\,\text{m/s}$), the baseline INDI controller achieves high-fidelity tracking. 
\\
Fig.~\ref{fig:states_1} details the time evolution of the system states in \eqref{eq:states} over a full trajectory lap, showing position (top), translational velocity (middle), and orientation (bottom) in solid blue. Under this nominal operating regime, the baseline controller $\boldsymbol{u}$ achieves near-zero tracking drift relative to the reference trajectory (dashed red).

\subsection{Spatio-Temporal Filter Performance} \label{subsec:B}
\begin{figure}[!b]
\centering
\includegraphics[width=0.485\textwidth]{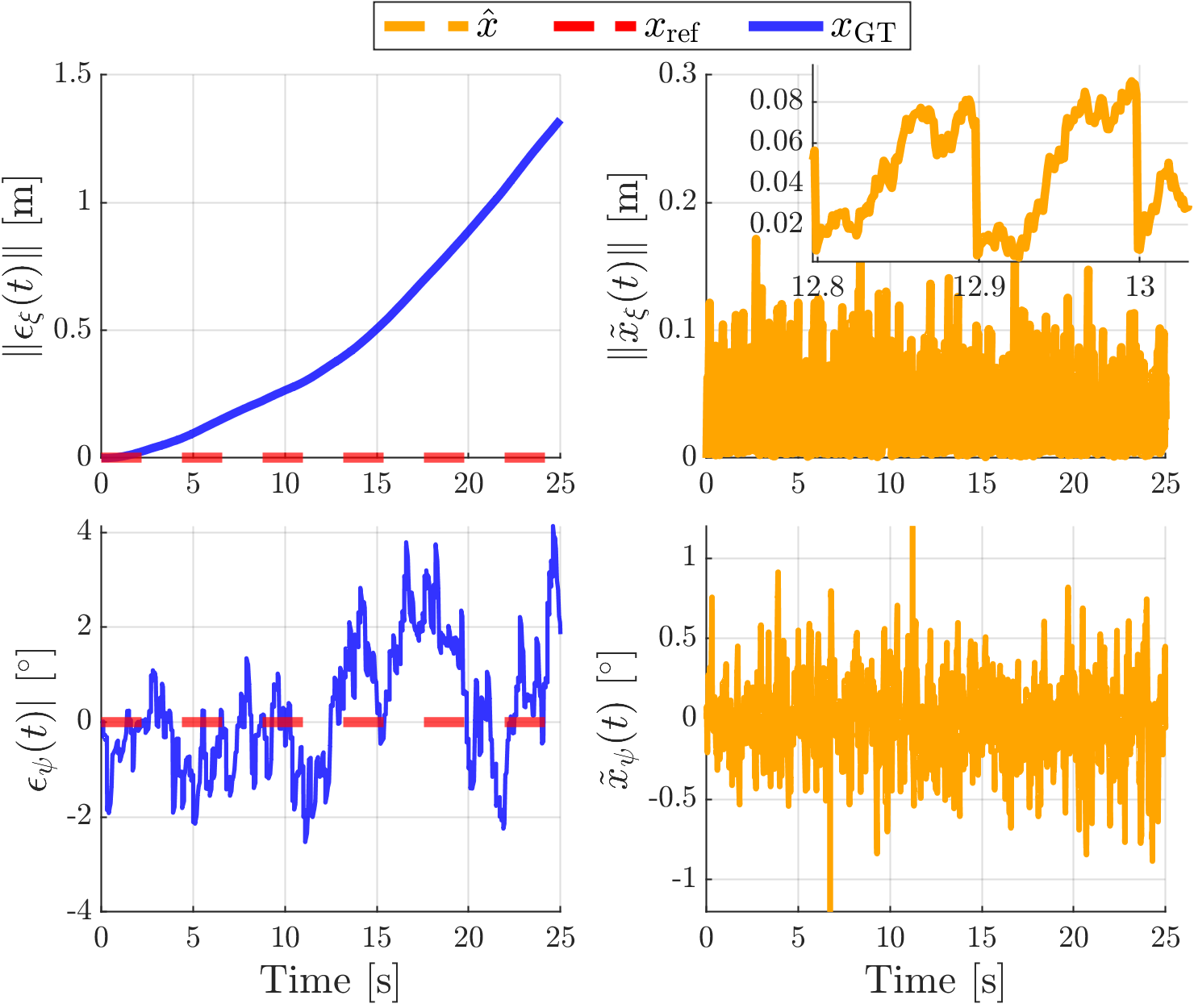}
\caption{Flat output tracking and estimation performance.}
\label{fig:statistical_moments}
\end{figure}

As our proposed EA mechanism directly depends on the filter's spatio-temporal fidelity, evaluating its space and time dynamics is critical. Formally, measurement innovations update the filter exclusively along covectors spanning the local observability codistribution $\Omega_{\mathrm{NL}}(\boldsymbol{x}) \subset T^*_{\boldsymbol{x}}\mathcal{M}$ \eqref{eq:codistribution}. Under smooth, low-acceleration flight, the rank condition of $\mathcal{O}_{\mathrm{NL}}(\boldsymbol{x})$ degenerates along weakly observed directions, allowing process noise propagation to dominate and inflating spatial uncertainty. Conversely, high-agility maneuvers, such as sharp transitions across the lemniscate intersection, provide rich kinematic excitation that expands $\text{span}(\Omega_{\mathrm{NL}}(\boldsymbol{x}))$, sharpening innovation updates and constraining error growth \cite{Hermann1977, Gauthier1994, martinelli2018nonlinear}.
\begin{figure}[!t]
\centering
\includegraphics[width=0.48\textwidth]{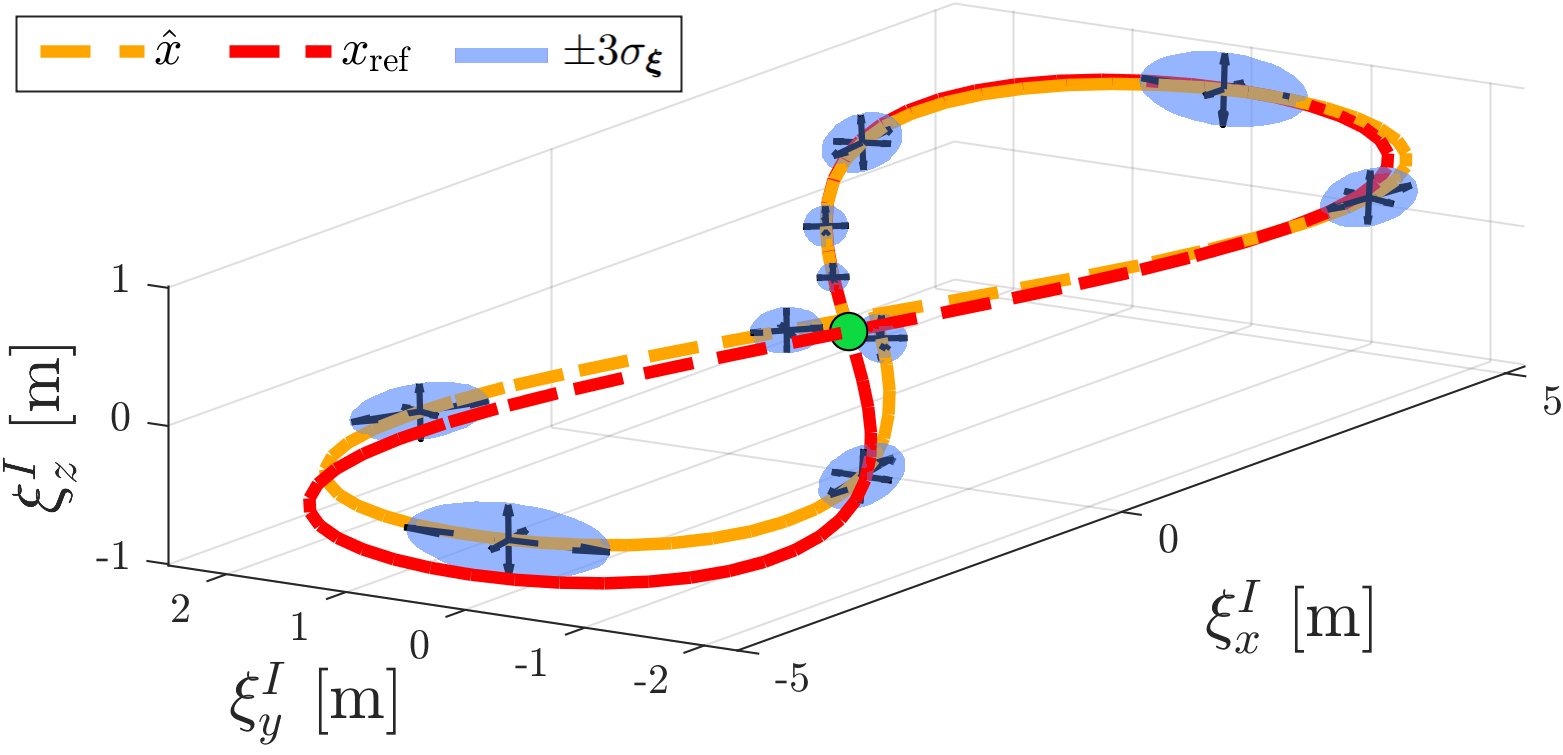}
\caption{Spatio-temporal covariance dynamics at $\mathrm{v}_{\mathrm{ref}} = 1.0\,\text{m/s}$, highlighting mid-air uncertainty expansion and contraction.}
\label{fig:flying_unc}
\end{figure}
Fig.~\ref{fig:flying_unc} visualizes these spatial dynamics via $3\sigma$ position error ellipsoids derived from the principal spectrum of $\mathbf{P}_{k\mid k}$ (pale blue), illustrating volume inflation and contraction driven by trajectory maneuvers. Fig.~\ref{fig:statistical_moments} complements this via the flat outputs ($\boldsymbol{\xi}^{\mathcal{I}}$ and $\psi$), contrasting tracking performance (left) against estimation errors (right). Despite stable 10~Hz filter updates, tracking performance degrades to a residual error of $\Vert{}\boldsymbol{\epsilon}_{\xi}\Vert{} \approx 1$~m upon lap completion.

\subsection{Observability–Uncertainty Mechanics} \label{subsec:C}
To analyze the evolution of local state constructibility along the reference trajectory, we evaluate the nonlinear observability metrics introduced in Section~\ref{subsec:obs}.

\subsubsection{Gramian Eigenspectrum}
\begin{figure}[!b]
\centering
\includegraphics[width=0.49\textwidth]{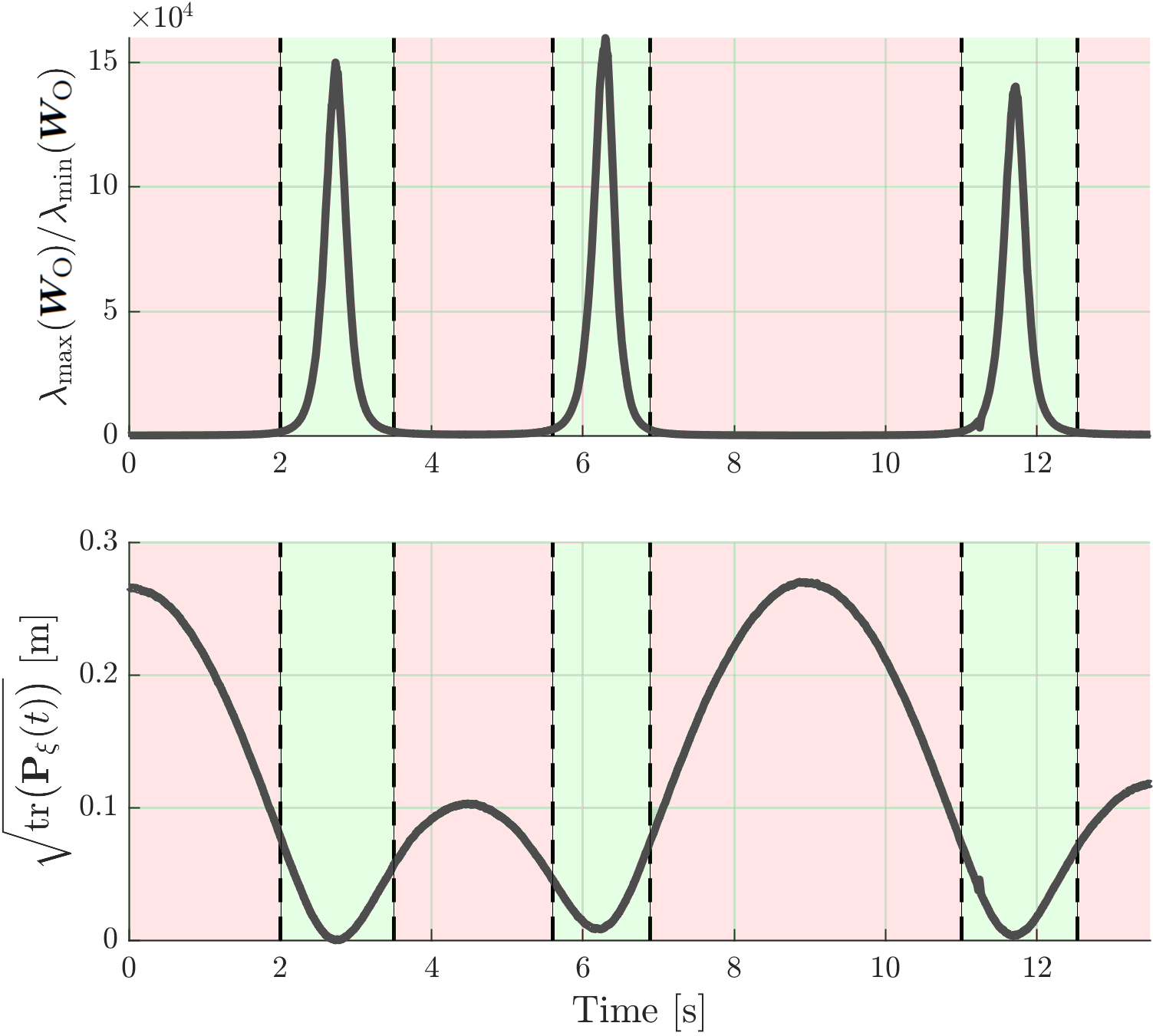}
\caption{Temporal interplay between condition number (top) and positional uncertainty (bottom) a half-lap orbit ($T/2$).}
\label{fig:obs_analysis}
\end{figure}

Temporal tracking pace directly governs estimator excitation: aggressive maneuvering along spatially constrained paths induces pronounced $\mathrm{SO}(3)$ attitude coupling and cross-accelerations. This dynamic excitation improves the local observability Gramian $\boldsymbol{W}_{\mathrm{O}}$, elevating its minimum eigenvalue $\lambda_{\min}(\boldsymbol{W}_{\mathrm{O}})$ along worst-case state directions. Consequently, as information throughput increases, the position error uncertainty $\sqrt{\mathrm{tr}(\mathbf{P}_{\boldsymbol{\xi}}(t))}$ contracts rapidly.
Conversely, straight-line segments relieve dynamic excitation, approaching low-acceleration regimes ($\ddot{\boldsymbol{\xi}}^{\mathcal{I}} \approx \mathbf{0}$, $\boldsymbol{\omega}^{\mathcal{B}} \approx \mathbf{0}$) where persistent excitation is lost. This degradation induces severe Gramian ill-conditioning $\kappa(\boldsymbol{W}_{\mathrm{O}})$~\eqref{eq:condition}, attenuating measurement updates and allowing weakly observable modes to drift.
\\
Fig.~\ref{fig:obs_analysis} illustrates this interplay over a half-lap ($T/2$) orbit. While condition number (top) spikes reflect instantaneous loss of excitation, their effect on positional uncertainty (bottom) exhibits a distinct phase lag. Because $\mathbf{P}(t)$ integrates process noise over time, uncertainty continues to inflate throughout unexcited flight, reaching its peak during extended straight segments (green regions). Dynamic maneuvers ultimately restore Gramian conditioning ($\kappa \to 1$), enabling measurement innovations to crush the uncertainty trace ($\sqrt{\mathrm{tr}(\mathbf{P}_{\boldsymbol{\xi}})} \downarrow$) down toward zero.

\subsubsection{EA Control Synthesis} 
\begin{figure}[!b] 
\centering 
\begin{tikzpicture}[
    >=Stealth,
    font=\sffamily\tiny,
    cellbox/.style={line width=1.2pt, rectangle, minimum width=3.6cm, minimum height=3.3cm, inner sep=0pt},
    vecFF/.style={-{Stealth[length=5pt, width=5pt]}, densely dashed, line width=2.0pt, cyan!95!white},
    vecFBx/.style={-{Stealth[length=5pt, width=5pt]}, densely dashed, line width=2.0pt, orange!90!white},
    vecFBe/.style={-{Stealth[length=5pt, width=5pt]}, densely dashed, line width=2.0pt, green!60!black},
    vecRes/.style={-{Stealth[length=7pt, width=8]}, double, line width=1.2pt, gold},
]

  \node[cellbox] (cell1) at (0,0) {};
  \node[cellbox] (cell2) at (4.1,0) {};
  \node[cellbox] (cell3) at (0,-3.8) {};
  \node[cellbox] (cell4) at (4.1,-3.8) {};

  \begin{scope}[shift={(cell1.center)}]
    \path let \p1 = ($(cell1.north east)-(cell1.south west)$) in
    node[anchor=center, inner sep=0pt, opacity=0.8] at (0,0) {%
      \includegraphics[width=\x1, height=\y1]{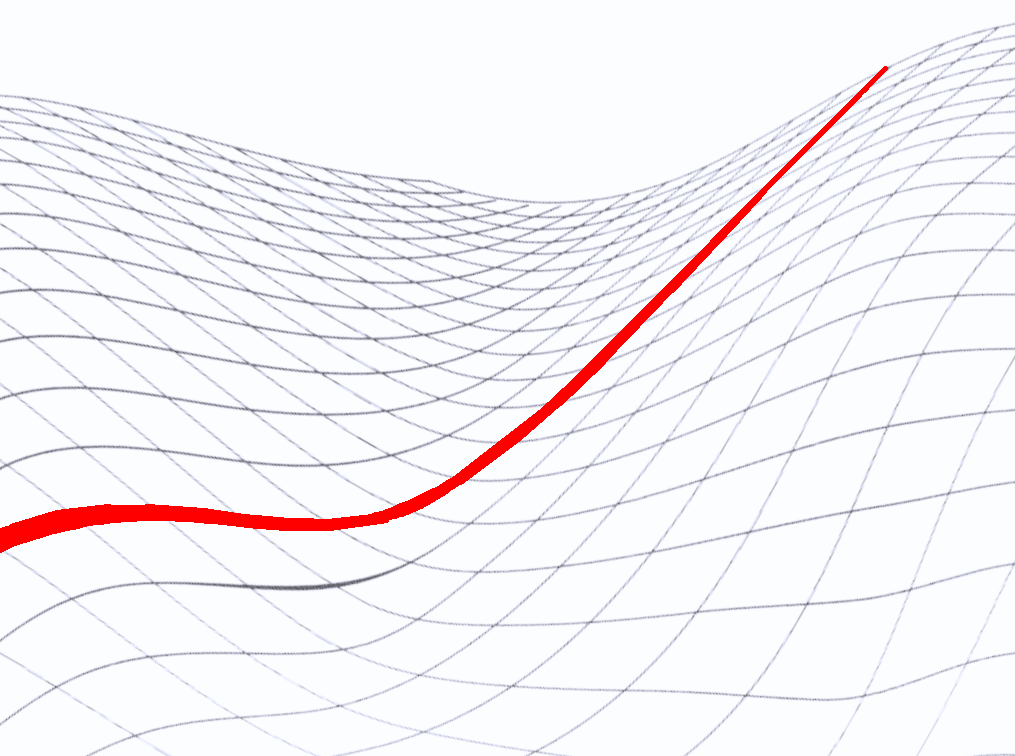}%
    };
    
    \node[anchor=north, font=\large] at (-1.55, 1.7) {($i$)};
    
    \node[anchor=north, font=\large] at (-0.0, 2.3) {$\boldsymbol{\epsilon}_{ \boldsymbol{y},k} = \boldsymbol{0}$};
    
    \node[anchor=south, rotate=90, font=\large] at (-1.9, 0) {$\boldsymbol{\delta} \hat{\boldsymbol{x}}_k = \boldsymbol{0}$};
    
    \coordinate (O1) at (-1.0, -0.6);
    \draw[black, double, line width=1.5pt, line cap=round] 
    ($(O1)+(-6pt,-3pt)$) -- ($(O1)+(6pt,3pt)$)
    ($(O1)+(-4pt,4pt)$)  -- ($(O1)+(4pt,-4pt)$);

    \draw[vecFF]  (O1) -- ++(45:1.0) coordinate (FF1);
    \coordinate (R1) at ($(FF1) $);
    \draw[vecRes] (O1) -- (R1);
    \draw[vecFF] (O1) -- ++(45:1.0) coordinate (FF1);

    \draw[black, line width=0.8pt] (cell1.south west) rectangle (cell1.north east);
  \end{scope}

  \begin{scope}[shift={(cell2.center)}]
    \path let \p1 = ($(cell1.north east)-(cell1.south west)$) in
    node[anchor=center, inner sep=0pt, opacity=0.8] at (0,0) {%
      \includegraphics[width=\x1, height=\y1]{Photos/Img_Grid_Top.png}%
      };
    \node[anchor=north, font=\large] at (-1.45, 1.7) {($ii$)};

    \node[anchor=north, font=\large] at (-0.0, 2.3) {$\boldsymbol{\epsilon}_{ \boldsymbol{y},k} \neq \boldsymbol{0}$};

    \coordinate (O2) at (-1.0, 0.2);
    \draw[black, double, line width=1.5pt, line cap=round] 
    ($(O2)+(-6pt,-3pt)$) -- ($(O2)+(6pt,3pt)$)
    ($(O2)+(-4pt,4pt)$)  -- ($(O2)+(4pt,-4pt)$);

    \draw[vecFF] (O2) -- ++(30:1.0) coordinate (FF2);
    \draw[vecFBe] (O2) -- ++(-85:1.0) coordinate (FBe2);
    \coordinate (R2) at ($(FF2) + (FBe2) - (O2)$);
    \draw[vecRes] (O2) -- (R2);
        
    \draw[black, line width=0.8pt] (cell2.south west) rectangle (cell2.north east);
  \end{scope}

  \begin{scope}[shift={(cell3.center)}]
    \path let \p1 = ($(cell1.north east)-(cell1.south west)$) in
    node[anchor=center, inner sep=0pt, opacity=0.8] at (0,0) {%
      \includegraphics[width=\x1, height=\y1]{Photos/Img_Grid_Full.png}%
    };
    \node[anchor=north, font=\large] at (-1.4, 1.7) {($iii$)};

    \node[anchor=south, rotate=90, font=\large] at (-1.9, 0) {$\boldsymbol{\delta} \hat{\boldsymbol{x}}_k \neq \boldsymbol{0}$};

    \coordinate (O3) at (-1.0, -0.6);
    \draw[black, double, line width=1.5pt, line cap=round] 
    ($(O3)+(-3pt,-7pt)$) -- ($(O3)+(3pt,7pt)$)
    ($(O3)+(-6pt,2pt)$)  -- ($(O3)+(6pt,-2pt)$);

    \draw[vecFF]  (O3) -- ++(45:1.0) coordinate (FF3);
    \draw[vecFBx] (O3) -- ++(95:0.9) coordinate (FBx3);
    \coordinate (R3) at ($(O3) + ($(FF3)-(O3)$) + ($(FBx3)-(O3)$) + ($(77.5:-0.3)$)$);
    \draw[vecRes] (O3) -- (R3);
    
    \draw[black, line width=0.8pt] (cell3.south west) rectangle (cell3.north east);
  \end{scope}

  \begin{scope}[shift={(cell4.center)}]
    \path let \p1 = ($(cell1.north east)-(cell1.south west)$) in
    node[anchor=center, inner sep=0pt, opacity=0.8] at (0,0) {%
      \includegraphics[width=\x1, height=\y1]{Photos/Img_Grid_Top.png}%
    };
    \node[anchor=north, font=\large] at (-1.45, 1.7) {($iv$)};
    
    \coordinate (O4) at (-1.0, 0.2);
    \draw[black, double, line width=1.5pt, line cap=round] 
    ($(O4)+(-3pt,-7pt)$) -- ($(O4)+(3pt,7pt)$)
    ($(O4)+(-6pt,2pt)$)  -- ($(O4)+(6pt,-2pt)$);

    \draw[vecFF]  (O4) -- ++(30:1.0) coordinate (FF4);
    \draw[vecFBe] (O4) -- ++(-85:0.8) coordinate (FBe4);
    \draw[vecFBx] (O4) -- ++(95:0.9) coordinate (FBx4);
    \coordinate (R4) at ($(O4) + ($(FF4)-(O4)$) + ($(FBe4)-(O4)$) + ($(FBx4)-(O4)$) + ($(-60:0.6)$)$);
    \draw[vecRes] (O4) -- (R4);
    
    \draw[black, line width=0.8pt] (cell4.south west) rectangle (cell4.north east);
  \end{scope}

\end{tikzpicture}
\caption{Vector decomposition of the proposed EA control increment \eqref{eq:u_increment} across tracking error ($\boldsymbol{\epsilon}_{\boldsymbol{y},k}$) and rate perturbation ($\boldsymbol{\delta}\hat{\boldsymbol{x}}_k$) regimes on the state manifold $\mathcal{M}$.}
\label{fig:confusion}
\end{figure}

Substituting the estimation-awareness factor \eqref{eq:awareness} into the baseline INDI control law \eqref{eq:u_EA} yields the augmented virtual control increment
\begin{align} \label{eq:u_increment}
\Delta \hat{\boldsymbol{\nu}}_k = \underbrace{\boldsymbol{y}_{\mathrm{ref},k+r}\vphantom{\sum\nolimits_{j=0}^{r-1}} }_{\mathrm{FF}} - \underbrace{ \sum\nolimits_{j=0}^{r-1} \mathbf{K}_j \boldsymbol{\epsilon}_{\boldsymbol{y},k+j} }_{\mathrm{FB, outer}} - \underbrace{ \overline{\mathbf{P}}_k \, \boldsymbol{\delta} \hat{\boldsymbol{x}}_k }_{\mathrm{FB, inner}} \, .
\end{align}
Guided by the error bounds established in Corollary~\ref{cor:tube}, this parameterization dynamically throttles feedback authority along weakly observed state directions whenever estimator confidence degrades.
Fig.~\ref{fig:confusion} illustrates this interaction as the quadrotor ($\boldsymbol{\times}$) tracks the reference path $\boldsymbol{y}_{\mathrm{ref}}$ (red). The resulting command $\Delta \hat{\boldsymbol{\nu}}_k$ (yellow) synthesizes predictive feedforward ($\mathrm{FF}$, cyan) with outer-loop spatial feedback ($\mathrm{FB}_{\text{outer}}$, green) and inner-loop feedback ($\mathrm{FB}_{\text{inner}}$, orange). Following the structural decomposition in Remark~\ref{rem:geometric_decomp}, we categorize this control geometry across four operational regimes governed by the spatial error $\boldsymbol{\epsilon}_{\boldsymbol{y},k}$ and state increment $\boldsymbol{\delta}\hat{\boldsymbol{x}}_k$:
\\
\textbf{(i) Pure Feedforward Mode ($\boldsymbol{\epsilon}_{\boldsymbol{y},k} = \mathbf{0}, \boldsymbol{\delta}\hat{\boldsymbol{x}}_k = \mathbf{0}$):} 
Nominal steady-state cruise. With zero error and zero perturbations, both feedback loops deactivate. The command reduces strictly to $\Delta \hat{\boldsymbol{\nu}}_k = \mathrm{FF}_k \parallel \mathbf{t}_{k+r}$, steering into upcoming path curvature via pure predictive drive.
\\
\textbf{(ii) Pure Kinematic Restoration ($\boldsymbol{\epsilon}_{\boldsymbol{y},k} \neq \mathbf{0}, \boldsymbol{\delta}\hat{\boldsymbol{x}}_k = \mathbf{0}$):} 
Pure spatial error correction. Inner rate damping remains off, while outer tracking feedback generates a transverse restoring vector $\mathrm{FB}_{\mathrm{outer}, k} \perp \mathbf{t}_{k}$. The vector sum rotates $\Delta \hat{\boldsymbol{\nu}}_k$ toward the path to restore tracking.
\\
\textbf{(iii) Dynamic Rate Corruption ($\boldsymbol{\epsilon}_{\boldsymbol{y},k} = \mathbf{0}, \boldsymbol{\delta}\hat{\boldsymbol{x}}_k \neq \mathbf{0}$):} 
Un-aided rate disturbance. Despite zero spatial error, unmitigated rate feedback generates a parasitic damping vector $\mathrm{FB}_{\mathrm{inner},k} = \boldsymbol{\delta}\hat{\boldsymbol{x}}_k$ (orange). This corrupts the orthogonal alignment, deflecting the resultant command $\Delta \hat{\boldsymbol{\nu}}_k$ off-axis away from the path tangent $\mathbf{t}_{k+r}$.
\\
\textbf{(iv) Estimation-Aware Co-Design ($\boldsymbol{\epsilon}_{\boldsymbol{y},k} \neq \mathbf{0}, \boldsymbol{\delta}\hat{\boldsymbol{x}}_k \neq \mathbf{0}$):} 
Simultaneous spatial recovery and rate mitigation. The correlation operator $\overline{\mathbf{P}}_k$ dynamically modulates the inner loop, suppressing uncoordinated jitter while allowing normal feedback $\mathrm{FB}_{\mathrm{outer},k} \perp \mathbf{t}_k$ to rotate the net command $\Delta \hat{\boldsymbol{\nu}}_k$ sharply back toward $\boldsymbol{y}_{\mathrm{ref}}$ for safe manifold re-entry.
\begin{figure*}[t] 
\centering
\includegraphics[width=1.0\textwidth]{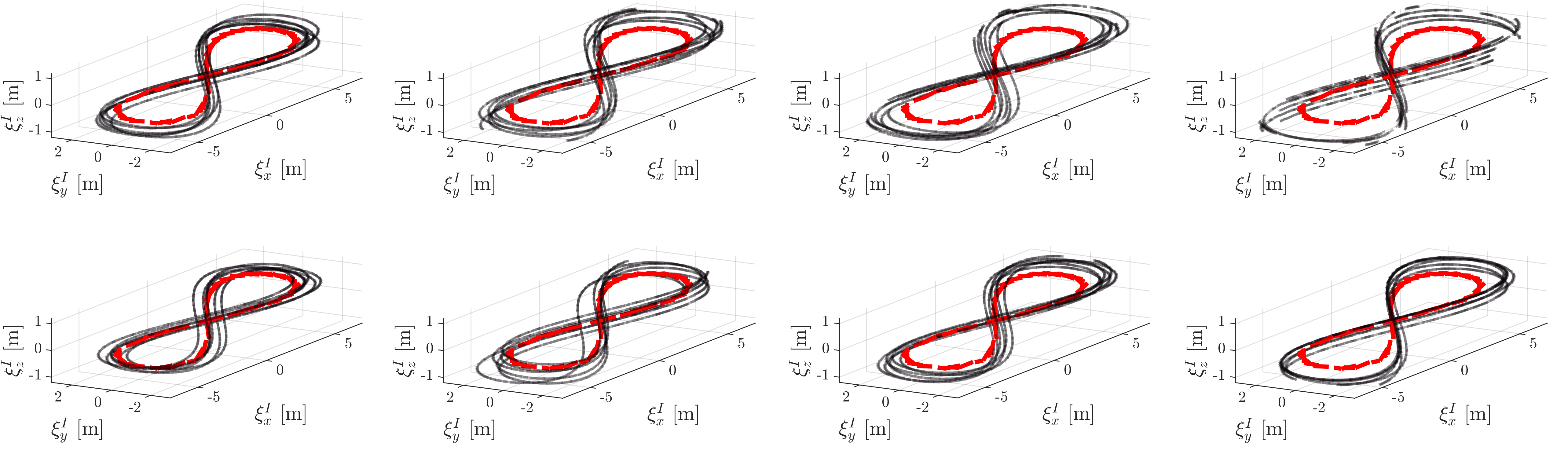}
\caption{Cross-controller benchmark across airspeed variations $\mathrm{v}_{\mathrm{ref}} \in \{1, 4, 7, 10\}\,\text{m/s}$ (left to right). Top row: unaugmented baseline control paradigms from Table~\ref{t:paradigms}; bottom row: respective EA-augmented formulations preserving trajectory fidelity.}
\label{fig:comparative}
\end{figure*}

\begin{table*}[!b]
\centering
\scriptsize
\caption{Performance summary across six control paradigms, reference airspeeds $\mathrm{v}_{\mathrm{ref}}$, and control regimes: unaided baselines (pale orange) vs. EA-augmented variants (pale blue). Relative percentage improvements are highlighted below (yellow, \textbf{\%}).}
\renewcommand{\arraystretch}{1.5} \setlength{\tabcolsep}{4.6pt}
\begin{tabular}{|c||c|c||c|c||c|c||c|c||c|c||c|c|}
\hline 
\multirow{2}{*}{\diagbox{$\mathrm{v}_{\mathrm{ref}}$}{$\boldsymbol{u}_j$}} & 
\multicolumn{2}{c|}{$\boldsymbol{u}_{\mathrm{FL}}$ \CC} &
\multicolumn{2}{c|}{$\boldsymbol{u}_{\mathrm{GC}}$ \CC} &
\multicolumn{2}{c|}{$\boldsymbol{u}_{\mathrm{CC}}$ \CC} &
\multicolumn{2}{c|}{$\boldsymbol{u}_{\mathrm{DF}}$ \CC} &
\multicolumn{2}{c|}{$\boldsymbol{u}_{\mathrm{NMPC}}$ \CC} &
\multicolumn{2}{c|}{$\boldsymbol{u}_{\mathrm{INDI}}$ \CC} \\ \cline{2-13}
& $\| \tilde{\boldsymbol{x}}_{\boldsymbol{\xi}} \| \,$[m] & $\| \boldsymbol{\epsilon}_{\boldsymbol{\xi}} \| \,$[m] & $\| \tilde{\boldsymbol{x}}_{\boldsymbol{\xi}} \| \,$[m] & $\| \boldsymbol{\epsilon}_{\boldsymbol{\xi}} \| \,$[m] & $\| \tilde{\boldsymbol{x}}_{\boldsymbol{\xi}} \| \,$[m] & $\| \boldsymbol{\epsilon}_{\boldsymbol{\xi}} \| \,$[m] & $\| \tilde{\boldsymbol{x}}_{\boldsymbol{\xi}} \| \,$[m] & $\| \boldsymbol{\epsilon}_{\boldsymbol{\xi}} \| \,$[m] & $\| \tilde{\boldsymbol{x}}_{\boldsymbol{\xi}} \| \,$[m] & $\| \boldsymbol{\epsilon}_{\boldsymbol{\xi}} \| \,$[m] & $\| \tilde{\boldsymbol{x}}_{\boldsymbol{\xi}} \| \,$[m] & $\| \boldsymbol{\epsilon}_{\boldsymbol{\xi}} \| \,$[m] \\ \hline \hline  
\multirow{3}{*}{\shortstack{{1.0 m/s} \\ {(\textit{3.6 km/h})}}} 
 & 0.045 \CO & 0.194 \CO & 0.078 \CO & 0.305 \CO & 0.072 \CO & 0.207 \CO & 0.069 \CO & 0.229 \CO & 0.044 \CO & 0.180 \CO & 0.036 \CO & 0.130 \CO \\ \cline{2-13}
 & 0.044 \CB & 0.192 \CB & 0.077 \CB & 0.310 \CB & 0.071 \CB & 0.210 \CB & 0.067 \CB & 0.227 \CB & 0.045 \CB & 0.184 \CB & 0.036 \CB & 0.129 \CB \\ \cline{2-13}
 & {2.21\%} \CY & {1.13\%} \CY & {1.21\%} \CY & {-1.79\%} \CY & {2.04\%} \CY & {-1.55\%} \CY & \textbf{3.32\%} \cellcolor{yellow} & {0.79\%} \CY & {-2.14\%} \CY & {-2.03\%} \CY & {1.02\%} \CY & {0.94\%} \CY \\ \hline \hline
\multirow{3}{*}{\shortstack{{4.0 m/s} \\ {(\textit{14.4 km/h})}}} 
 & 0.177 \CO & 0.380 \CO & 0.243 \CO & 0.525 \CO & 0.150 \CO & 0.522 \CO & 0.132 \CO & 0.559 \CO & 0.113 \CO & 0.419 \CO & 0.086 \CO & 0.245 \CO \\ \cline{2-13}
 & 0.162 \CB & 0.364 \CB & 0.226 \CB & 0.507 \CB & 0.138 \CB & 0.499 \CB & 0.122 \CB & 0.531 \CB & 0.102 \CB & 0.396 \CB & 0.074 \CB & 0.230 \CB \\ \cline{2-13}
 & {8.13\%} \CY & {4.12\%} \CY & {6.77\%} \CY & {3.46\%} \CY & {8.03\%} \CY & {4.52\%} \CY & {7.31\%} \CY & {5.12\%} \CY & {9.14\%} \CY & {5.33\%} \CY & \textbf{13.30\%} \cellcolor{yellow} & \textbf{6.11\%} \cellcolor{yellow} \\ \hline \hline 
\multirow{3}{*}{\shortstack{{7.0 m/s} \\ {(\textit{25.2 km/h})}}} 
 & 0.389 \CO & 0.781 \CO & 0.449 \CO & 0.832 \CO & 0.274 \CO & 1.052 \CO & 0.328 \CO & 1.021 \CO & 0.245 \CO & 0.827 \CO & 0.227 \CO & 0.590 \CO \\ \cline{2-13}
 & 0.304 \CB & 0.643 \CB & 0.383 \CB & 0.745 \CB & 0.228 \CB & 0.940 \CB & 0.279 \CB & 0.906 \CB & 0.207 \CB & 0.733 \CB & 0.167 \CB & 0.491 \CB \\ \cline{2-13}
 & {21.78\%} \CY & \textbf{17.77\%} \cellcolor{yellow} & {14.73\%} \CY & {10.48\%} \CY & {16.67\%} \CY & {10.64\%} \CY & {14.93\%} \CY & {11.25\%} \CY & {15.33\%} \CY & {11.41\%} \CY & \textbf{26.74\%} \cellcolor{yellow} & {16.68\%} \CY \\ \hline \hline
%
\multirow{3}{*}{\shortstack{{10.0 m/s} \\ {(\textit{36.0 km/h})}}} 
 & 0.569 \CO & 1.203 \CO & 0.885 \CO & 1.233 \CO & 0.464 \CO & 1.588 \CO & 0.550 \CO & 1.778 \CO & 0.516 \CO & 1.429 \CO & 0.407 \CO & 1.108 \CO \\ \cline{2-13}
 & 0.436 \CB & 0.952 \CB & 0.650 \CB & 1.048 \CB & 0.366 \CB & 1.324 \CB & 0.442 \CB & 1.438 \CB & 0.378 \CB & 1.174 \CB & 0.272 \CB & 0.838 \CB \\ \cline{2-13}
 & {23.37\%} \CY & {20.88\%} \CY & {26.54\%} \CY & {14.99\%} \CY & {21.13\%} \CY & {16.60\%} \CY & {19.67\%} \CY & {19.14\%} \CY & {26.78\%} \CY & {17.83\%} \CY & \textbf{33.24\%} \cellcolor{yellow} & \textbf{24.34\%} \cellcolor{yellow} \\ \hline
\end{tabular} \label{t:comparative}
\end{table*}

\subsection{Comparative Analysis} \label{subsec:D}
To evaluate the efficacy and generality of the proposed EA mechanism, we benchmark all six control paradigms listed in Table~\ref{t:paradigms} along the high-agility 3D lemniscate trajectory across two operational axes:
\begin{enumerate}[label=\roman*), itemsep=1.5pt, parsep=1pt, topsep=2pt]
\item \textbf{Flight Speed} (Left to Right): Assessing tracking robustness as reference airspeed increases from mild cruising ($\mathrm{v}_{\mathrm{ref}} = 1.0\,\text{m/s}$) to high-agility flight ($\mathrm{v}_{\mathrm{ref}} = 10.0\,\text{m/s}$), where aerodynamic forces dominate quadrotor dynamics.
\item \textbf{Estimation-Awareness} (Top vs. Bottom): Evaluating whether integrating the EA mechanism improves state-tracking fidelity across all baseline architectures.
\end{enumerate}
As shown in Fig.~\ref{fig:comparative}, the trajectory bundles reveal a pronounced speed-dependent divergence from left to right. In the unaugmented implementations (top row, solid gray), increasing flight speed progressively amplifies phase lag, high-frequency path hunting, and spatial drift away from the reference trajectory ($\boldsymbol{x}_{\mathrm{ref}}$, dashed red). Conversely, the EA-augmented variants (bottom row) maintain tight, coherent trajectory containment around the reference path across the entire velocity spectrum.
\\
Table~\ref{t:comparative} quantifies these visual trends, detailing performance over the $T$-second flight profile via the root mean square error (RMSE): $\| \cdot \| \triangleq \sqrt{ \mathbb{E}_k \, [ \, \left\| ( \, \cdot \, )_{k} \right\|^2 ] } $, where $(\cdot)_k$ evaluates either the estimation  $\tilde{\boldsymbol{x}}_{\boldsymbol{\xi},k}$ or tracking $\boldsymbol{\epsilon}_{\boldsymbol{\xi},k}$ error vectors.
In low-speed flight ($\mathrm{v}_{\mathrm{ref}} = 1\,\text{m/s}$), where estimates remain accurate and sensor noise dominates, all baseline models exhibit a nominal tracking RMSE of roughly $22 \pm 9\,\text{cm}$, rendering the EA effect negligible ($\approx \pm 2\%$). However, as reference airspeeds approach $10\,\text{m/s}$, the unaugmented baselines face severe structural difficulties, inflating the tracking RMSE to $1.1 \pm 0.2\,\text{m}$. Conversely, integrating the EA mechanism yields performance gains that scale rapidly from modest single-digit reductions at low speeds to substantial double-digit improvements at elevated airspeeds across all paradigms.
Notably, the EA-augmented INDI architecture ($\boldsymbol{u}_{\mathrm{INDI}}$) almost consistently achieves the highest metrics—reaching peak estimation and tracking error reductions of $33.24\%$ and $24.34\%$, respectively, at maximum speed. This uniform performance recovery confirms that the proposed EA mechanism provides a robust, universal mechanism for stabilizing estimator-controller interactions under transient estimation degradation.

\subsection{Discussion: Invariance and Robustness}
\begin{figure}[!h]
\centering
\includegraphics[width=0.49\textwidth]{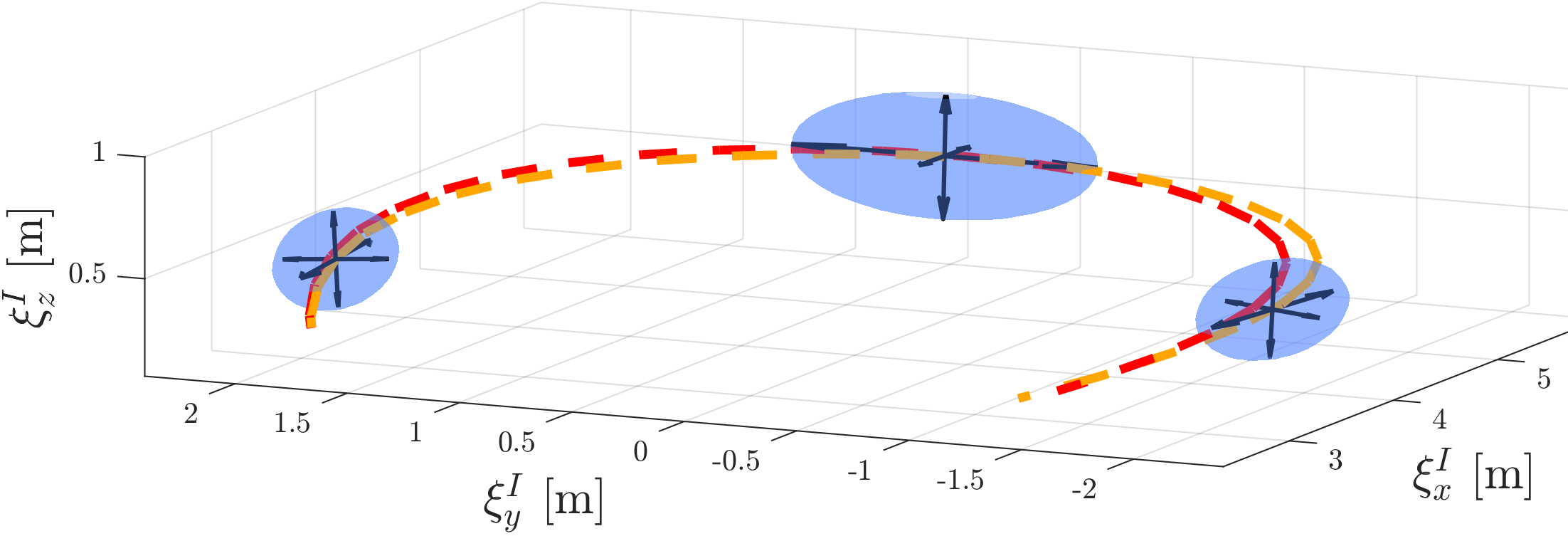}
\caption{Position error ellipsoids along the flight trajectory. The principal axes indicate the eigen-directions ($\boldsymbol{v}_i$) and magnitudes ($\sqrt{\lambda_i}$) of $\mathbf{P}_{k\mid k}$, which parameterize the anisotropic attenuation in the proposed EA control framework \cite{kay1993fundamentals, engelsman2023parametric}.}
\label{fig:ellipsoid}
\end{figure}

Formally, standard dual-loop control architectures process state estimates deterministically, operating under the implicit assumption of perfect estimator fidelity ($\hat{\boldsymbol{x}} \equiv \boldsymbol{x}$). However, during observability loss, the posterior error covariance $\mathbf{P}_{k\mid k}$ inflates toward its fundamental lower limit dictated by the inverse Fisher information matrix (FIM) \cite{SIMANDL20011703}, $\mathbf{P}_{k\mid k} \succeq \boldsymbol{\mathcal{I}}^{-1}(\boldsymbol{x}_k)$. As illustrated in Fig.~\ref{fig:ellipsoid}, this dynamic covariance expansion alters both the spatial orientation and magnitude of the uncertainty ellipsoids along the trajectory. Under conventional feedback laws, high-gain corrections yield spurious control inputs that inject parasitic energy into the dynamics, corrupting estimator innovations and expanding the closed-loop invariant tube radius $\gamma_k$ in \eqref{eq:Inv_tube}.
\\
Conversely, the proposed EA operator establishes an anisotropic feedback mapping that dynamically attenuates control authority along weakly observed state directions \cite{Barmish1983}, specifically targeting tangent vectors $\boldsymbol{v} \in T_{\boldsymbol{x}}\mathcal{M}$ associated with large covariance eigenvalues. By suppressing parasitic control excitation during transient sensing loss, the EA mechanism prevents artificial covariance amplification. This spectral filtering enforces a strictly tighter invariant tube bound, $\gamma_{k}^{\text{EA}} < \gamma_{k}^{\text{Base}}$, guaranteeing trajectory drift containment while constraining control effort near theoretical estimation limits.

\section{Conclusion} \label{sec:conc}
This paper presented a generalized EA control framework for quadrotor trajectory tracking, synthesized across multiple nonlinear control paradigms to resolve feedback vulnerability to estimator degradation. By explicitly coupling feedback authority to the posterior error covariance $\mathbf{P}_{k\mid k}$ via a normalized awareness operator $\overline{\mathbf{P}}_k$, the proposed architecture dynamically scales closed-loop gains according to real-time estimator confidence.
Our theoretical analysis and spectral analysis demonstrated that dynamic maneuvers boost local Gramian conditioning ($\kappa(\mathbf{W}_{\mathrm{O}}) \to 1$) by elevating $\lambda_{\min}(\mathbf{W}_{\mathrm{O}})$, driving rapid uncertainty contraction ($\sqrt{\mathrm{tr}(\mathbf{P}_{\boldsymbol{\xi}})} \downarrow$). In unexcited flight regimes where Gramian ill-conditioning induces transient covariance inflation, the EA mechanism acts as an anisotropic spectral filter along the tangent space $T_{\boldsymbol{x}}\mathcal{M}$. 
By suppressing parasitic control excitation along weakly observed directions, the EA framework prevents drift accumulation and enforces strictly tighter closed-loop invariant tube bounds. 
\\
Numerical simulations across diverse baseline controllers confirm that incorporating the EA framework significantly enhances tracking precision under severe state uncertainty and estimation degradation. Demonstrating these advantages, EA-augmented INDI yields $33.24\%$ and $24.34\%$ reductions across primary tracking metrics at high-speed flight ($10\text{~m/s}$). Building on these numerical results, future work will pursue experimental flight validation in agile acrobatic scenarios and obstacle-dense indoor spaces.

\bibliographystyle{elsarticle-num}
\bibliography{Ref}

\end{document}